\documentclass[10pt,english,compsoc,finalsubmission,twoside]{IEEEtran}
\usepackage[T1]{fontenc}
\usepackage[utf8]{inputenc}
\usepackage{array}
\usepackage{booktabs}
\usepackage{graphicx}

\makeatletter

\providecommand{\tabularnewline}{\\}

\usepackage[noadjust]{cite}
\usepackage{hyperref}
\usepackage{rotating}
\usepackage{flushend}

\makeatother

\usepackage{babel}

\begin{document}
\title{Evaluation and Measurement of Software Process Improvement - A Systematic Literature Review}
\author{Michael~Unterkalmsteiner,~\IEEEmembership{Student Member,~IEEE,} Tony~Gorschek,~\IEEEmembership{Member,~IEEE,} A.\,K.\,M.~Moinul~Islam, Chow~Kian~Cheng, Rahadian~Bayu~Permadi, Robert~Feldt,~\IEEEmembership{Member,~IEEE}
\IEEEcompsocitemizethanks{\IEEEcompsocthanksitem M.~Unterkalmsteiner, T.~Gorschek and R.~Feldt are with the Software Engineering Research Lab, School of Computing, Blekinge Institute of Technology, SE-371~79 Karlskrona, Sweden. E-mail: \{mun, tgo, rfd\}@bth.se.
\IEEEcompsocthanksitem A.\,K.\,M.~Moinul Islam is with the Software Engineering: Process and Measurement Research Group, Department of Computer Science, University of Kaiserslautern, PO Box 3049, 67653 Kaiserslautern, Germany. E-mail: moinul.islam@cs.uni-kl.de.
\IEEEcompsocthanksitem C.\,K.~Cheng is with General Electrics Healthcare, Healthcare IT, Munzinger Straße 5, 79111 Freiburg, Germany. E-mail: ChowKian.Cheng@ge.com.
\IEEEcompsocthanksitem R.\,B.~Permadi is with Amadeus S.\,A.\,S., Product Marketing and Development, 485 Route du Pin Montard, Boite Postale 69, 06902 Sophia Antipolis Cedex, France. E-mail: rahadian-bayu.permadi@amadeus.com.}
\thanks{Manuscript received May 11, 2010; revised February 8, 2011; accepted February 15, 2011}}
\IEEEcompsoctitleabstractindextext{%
\begin{abstract}%
BACKGROUND---Software Process Improvement (SPI) is a systematic approach to increase the efficiency and effectiveness of a software development organization and to enhance software products.
OBJECTIVE---This paper aims to identify and characterize evaluation strategies and measurements used to assess the impact of different SPI initiatives. 
METHOD---The systematic literature review includes 148 papers published between 1991 and 2008. The selected papers were classified according to SPI initiative, applied evaluation strategies and measurement perspectives. Potential confounding factors interfering with the evaluation of the improvement effort were assessed. 
RESULTS---Seven distinct evaluation strategies were identified, whereas the most common one, {}``Pre-Post Comparison'', was applied in 49\% of the inspected papers. Quality was the most measured attribute (62\%), followed by Cost (41\%) and Schedule (18\%). Looking at measurement perspectives, {}``Project'' represents the majority with 66\%. 
CONCLUSION---The evaluation validity of SPI initiatives is challenged by the scarce consideration of potential confounding factors, particularly given that {}``Pre-Post Comparison'' was identified as the most common evaluation strategy, and the inaccurate descriptions of the evaluation context. Measurements to assess the short and mid-term impact of SPI initiatives prevail, whereas long-term measurements in terms of customer satisfaction and return on investment tend to be less used.
\end{abstract}}
\maketitle

\section{Introduction\label{sec:Introduction}}

\IEEEPARstart{W}{ith} the increasing importance of software products
in industry as well as in our every day's life~\cite{kitchenham_software_1996},
the process of developing software has gained major attention by software
engineering researchers and practitioners in the last three decades~\cite{wirth_brief_2008,shaw_prospects_1990,scacchi_process_2001,sjoberg_future_2007}.
Software processes are human-centered activities and as such prone
to unexpected or undesired performance and behaviors~\cite{fuggetta_software_2000}.
It is generally accepted that software processes need to be continuously
assessed and improved in order to fulfill the requirements of the
customers and stakeholders of the organization~\cite{fuggetta_software_2000}.
Software Process Improvement (SPI) encompasses the assessment and
improvement of the processes and practices involved in software development~\cite{card_research_2004}.
\emph{SPI initiatives} are henceforth referred to activities aimed
at improving the software development process (see Section~\ref{sub:SPI-initiative}
for a definition of the different types of initiatives). 

The measurement of the software process is a substantial component
in the endeavor to reach predictable performance and high capability,
and to ensure that process artifacts meet their specified quality
requirements~\cite{paulish_case_1994,florac_measuring_1999}. As
such, software measurement is acknowledged as essential in the improvement
of software processes and products since, if the process (or the result)
is not measured and evaluated, the SPI effort could address the wrong
issue~\cite{hall_measurement_2001}. 

Software measurement is a necessary component of every SPI program
or change effort, and empirical results indicate that measurement
is an important factor for the initiatives' success~\cite{dyba_empirical_2005,goldenson_empirical_1996}.
The feedback gathered by software measurement and the evaluation of
the effects of the improvement provide at least two benefits. By making
the outcome visible, it motivates and justifies the effort put into
the initiative. Furthermore, it enables assessment of SPI strategies
and tactics~\cite{mathiassen_managing_2005}. However, at the same
time, it is difficult to establish and implement a measurement program
which provides relevant and valid information on which decisions can
be based~\cite{brown_measurement_2004,mathiassen_managing_2005}.
There is little agreement on what should be measured, and the absence
of a systematic and reliable measurement approach is regarded as a
factor that contributes to the high failure rate of improvement initiatives~\cite{iversen_problems_2006}.
Regardless of these problems in evaluating SPI initiatives, a plethora
of evidence exists to show that improvement efforts provide the expected
benefits~\cite{ferreira_roi_2008,mohagheghi_empirical_2008,redzic_six_2006,canfora_applying_2006,sommerville_empirical_2005,hyde_intangible_2004,goldenson_demonstrating_2003,achatz_industrial_2003,murugappan_blending_2003,trienekens_product_2001}. 

An interesting question that arises from that is how these benefits
are actually assessed. A similar question was raised by Gorschek and
Davis~\cite{gorschek_requirements_2008}, where it was criticized
how changes~/~improvements in requirements engineering processes
are evaluated for their success. Inspired by the search for dependent
variables~\cite{gorschek_requirements_2008}, we conducted a Systematic
Literature Review (SLR) to explore how the success of SPI initiatives
is determined, and if the approach is different depending on the particular
initiative. Furthermore, we investigated which types of measures are
used and, based on the categorization by Gorschek and Davis~\cite{gorschek_requirements_2008},
which perspectives (project, product or organization) are used to
assess improvement initiatives. Following the idea of Evidence-Based
Software Engineering (EBSE)~\cite{dyba_empirical_2005} we collect
and analyze knowledge from both research and practical experience.
To this end we adopted the approach for conducting SLR's proposed
by Kitchenham~\cite{kitchenham_guidelines_2007}. 

This paper is organized as follows. Background and related work is
presented in Section~\ref{sec:Background-and-Related} and our research
methodology is presented in Section~\ref{sec:Research-Methodology}.
In Section~\ref{sec:Results-and-Analysis} we describe the results
and answer our four major research questions. We present our conclusions
in Section~\ref{sec:Conclusion}.

\section{Background and Related Work\label{sec:Background-and-Related}}

\subsection{Software process improvement \label{sub:Software-process-improvement}}

Software process research is motivated by the common assumption that
process quality is directly related with the quality of the developed
software~\cite{kitchenham_software_1996,cugola_software_1998,fuggetta_software_2000}.
The aim of software process improvement is therefore to increase product
quality, but also to reduce time-to-market and production costs~\cite{cugola_software_1998}.
The mantra of many software process improvement frameworks and models
originates in the Shewhart-Deming cycle~\cite{deming_out_1986}:
establish an improvement plan, implement the new process, measure
the changed process, and analyze the effect of the implemented changes~\cite{humphrey_introduction_1993,fox_quality_1997,gorschek_packaging_2004,pettersson_practitioners_2008}.

The Capability Maturity Model (CMM)~\cite{paulk_capability_1995}
is an early attempt to guide organizations to increase their software
development capability and process maturity~\cite{boehm_view_2006}.
Although software and process measurement is an integral part of the
lower maturity levels (repeatable and defined) and central for the
managed level~\cite{paulk_capability_1993}, the model only \emph{suggests}
concrete measurements since the diversity of project environments
may evoke varying measurement needs~\cite{paulk_key_1993}. Similarly,
the Capability Maturity Model Integration (CMMI)~\cite{ahern_cmmism_2001,_capability_2002-1,_capability_2002}
and ISO\,/\,IEC~15504~\cite{el_emam_spice:_1998,_iso/iec_1998}
(also known as SPICE), propose various measurements. The CMMI reference
documentation, both for the staged and the continuous representation~\cite{_capability_2002-1,_capability_2002},
provides measurement suggestions for each process area as an informative
supplement to the required components of the model. The ISO\,/\,IEC~15504
standard documentation~\cite{_iso/iec_1998-1}, on the other hand,
prescribes that the process improvement has to be confirmed and defines
a process measurement framework. The informative part of the ISO standard
provides some rather limited examples of process measures without
showing how the measurement framework is applied in practice.

A common characteristic of the above-mentioned improvement initiatives
is their approach to identify the to-be-improved processes: the actual
processes are compared against a set of {}``best practice'' processes.
In case of significant divergences, improvement opportunities are
identified and the elimination of the differences constitutes the
actual process improvement~\cite{thomas_top-down_1994}. This approach
is commonly referred to as top-down~\cite{thomas_top-down_1994}
or prescriptive~\cite{pettersson_practitioners_2008} improvement.
In conceptual opposition to this idea are the bottom-up~\cite{thomas_top-down_1994}
or inductive~\cite{pettersson_practitioners_2008} approaches to
process improvement. The main principle of bottom-up improvement is
a process change driven by the knowledge of the development organization
and not by a set of generalized {}``best practices''~\cite{thomas_top-down_1994}.
The Quality Improvement Paradigm (QIP)~/~Experience Factory~\cite{basili_experience_1993,basili_improve_1995}
is one instance in this category of improvement initiatives. As in
the prescriptive approaches, measurement to control process change
and to confirm goal achievement is a central part of QIP.

\subsection{Related work}

Gorschek and Davis present a conceptual framework for assessing the
impact of requirements process changes~\cite{gorschek_requirements_2008}.
Their central idea is that the effect of a change in the requirements
process can be observed and measured at different levels: (1)~Effort
and quality of requirements related activities and artifacts in the
requirements phase, (2)~project success in terms of meeting time,
budget and scope constraints, (3)~product success in terms of meeting
both the customers' and the company's expectations, (4)~company success
in terms of product portfolio and market strategies, and (5)~the
influence on society.

Although these concepts are described from the perspective of requirements
engineering, the essence to evaluate a process change on different
levels to understand its impact more thoroughly, is conveyable to
software process improvement in general. 

By looking at the recent literature one can find several endeavors
to systematically collect and analyze the current knowledge in software
measurement.

Gomez et al.~\cite{gomez_systematic_2008} conducted a SLR on measurement
in software engineering. The study considered in total 78 publications
and tried to answer three questions: “What to measure?”, “How to Measure?’
and “When to Measure?” The criteria for inclusion in the review were
that the publication presents current and useful measurements. To
answer the first question, the study accumulated the metrics based
on entities where the measures are collected and the measured attributes.
The most measured entity was {}``Product'' (79\%), followed by {}``Project''
(12\%) and {}``Process'' (9\%), and the most measured attributes
were {}``Complexity'' (19\%), {}``Size'' (16\%) and {}``Inheritance''
(8\%). The second question is answered by identifying metrics that
have been validated empirically (46\%), theoretically (26\%) and both
empirically\,/\,theoretically (28\%). Furthermore the measurement
focus, e.\,g.\ object-orientation, process, quality, was analyzed.
The answer for the third question, when to measure, is presented by
mapping the metrics onto the waterfall lifecycle phases. The identified
product metrics are found in the design (42\%), development (27\%),
maintenance (14\%), testing (12\%) and analysis (5\%) phase. 

Bellini et al.~\cite{bellini_measurement_2008} systematically reviewed
the literature in twenty Software Engineering and Information Systems
journals with the aim of describing current and future trends in software
measurement. The study identifies and discusses five key software
measurement topics in the the reviewed literature: measurement theory,
software metrics, development and identification of metrics, measure
collection, and evaluation and analysis of measures. The authors conclude
that, besides traditional software measures like code complexity and
developer productivity, developments from organizational studies,
marketing and human resources management are gaining interest in the
area of Software Engineering\,/\,Information Systems due to the
human-intensive nature of software development. Measures used in practice
should be developed based upon a common agreement on the relationship
between the empirical object of interest and its mathematical representation.
Furthermore, for the practical analysis of measures, a more flexible
interpretation of the admissible transformations of measurement scales
is advocated.

Kitchenham~\cite{kitchenham_whats_2010} conducted a systematic mapping
study to describe the state-of-the-art in software metrics research.
The study assesses 103 papers published between 2000 and 2005 and
includes an analysis on their influence (in terms of citation counts)
on research. Kitchenham concludes that papers presenting empirical
validations of metrics have the highest impact on metrics research
although she has also identified several issues with this type of
studies. For example, 5 out of 7 papers, which empirically validated
the object oriented metrics proposed by Chidamber and Kemerer~\cite{chidamber_metrics_1994},
included Lack of Cohesion (LCOM) in the validation. Kitchenham~\cite{kitchenham_whats_2010}
pointed out that LCOM has been demonstrated theoretically invalid~\cite{hitz_chidamber_1996}
and that continuous attempts to validate LCOM empirically seem therefore
futile.

The aim of this SLR differs from the above reviews in two aspects.
First, the focus of this review is on measurement of software process
improvement initiatives, i.\,e.\ what to measure, and is therefore
more specific than the reviews of Bellini et al.\ and Gomez et al.
Second, this review investigates also how the measures are used to
evaluate and analyze the process improvement. Given our different
focus, only 1 (\cite{iversen_cultivation_2003}) of our 148 reviewed
papers was also covered by Bellini et al.~\cite{bellini_measurement_2008}.
Gomez et al.~\cite{gomez_systematic_2008} did not report the reviewed
papers which impedes a coverage assessment with our SLR.

\section{Research Methodology\label{sec:Research-Methodology}}

\begin{figure}
\begin{centering}
\includegraphics[height=30pc]{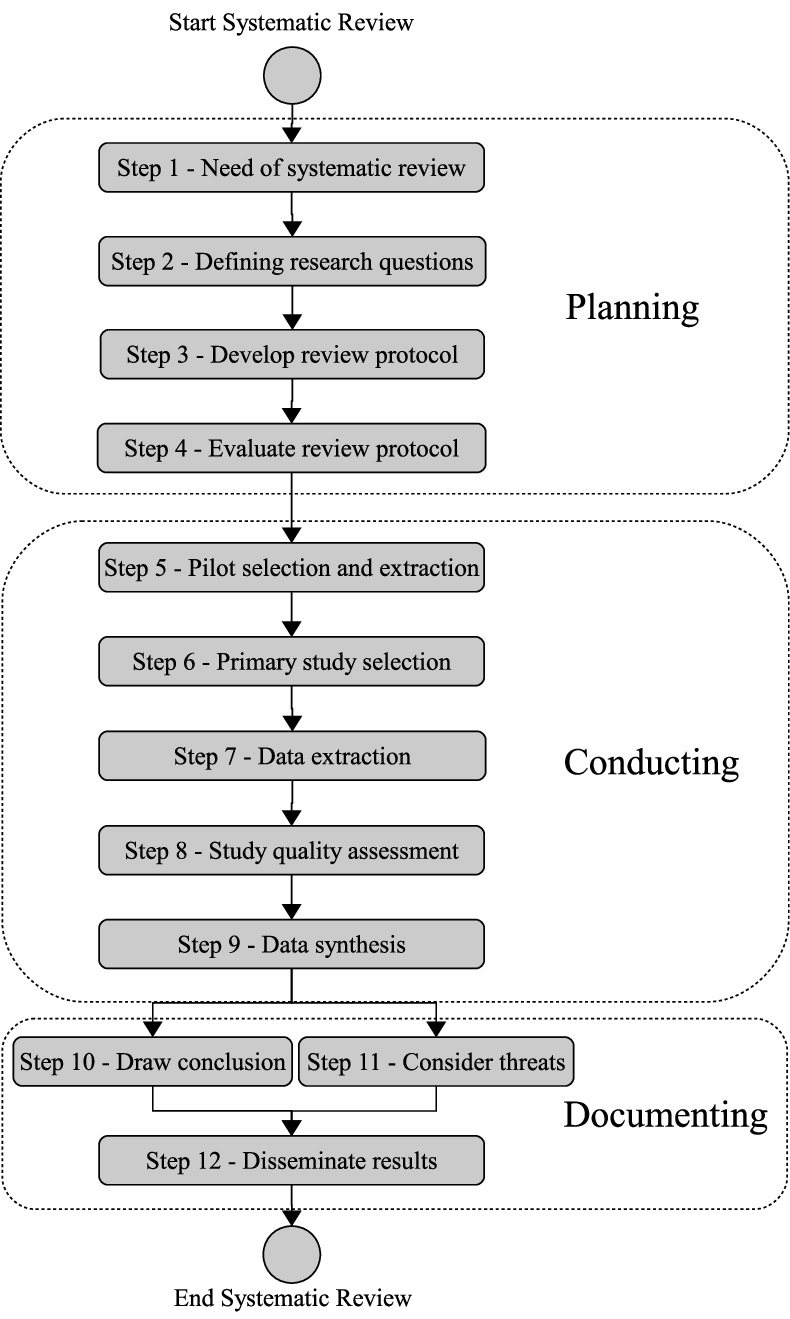}
\par\end{centering}

\caption{Systematic review steps (adapted from \cite{kitchenham_guidelines_2007})\label{fig:Systematic-review-steps}}
\end{figure}

In this section we describe the design and the execution of the SLR.
Furthermore, we discuss threats to the validity of this review. Figure~\ref{fig:Systematic-review-steps}
outlines the research process we have used and its steps are described
in detail in the following sub-sections.

\begin{table*}
\caption{Research questions for the systematic review\label{tab:Research-questions}}

\begin{tabular}{c>{\raggedright}p{3in}>{\raggedright}p{3.5in}}
\toprule 
ID & \multicolumn{1}{c}{Question} & \multicolumn{1}{c}{Aim}\tabularnewline
\midrule
\midrule 
RQ1 & What types of evaluation strategies are used to evaluate SPI initiatives? & To identify which concrete evaluation strategies are used and how
they are applied in practice to assess SPI initiatives. \tabularnewline
\midrule 
RQ2 & What are the reported metrics for evaluating the SPI initiatives? & To identify the metrics which are commonly collected and used to evaluate
SPI initiatives. \tabularnewline
\midrule 
RQ3 & What measurement perspectives are used in the evaluation and to what
extent are they associated with the identified SPI initiatives? & To determine from which measurement perspective SPI initiatives are
evaluated. Furthermore, to analyze any relationship between SPI initiatives
and measurement perspectives. \tabularnewline
\midrule 
RQ4 & What are the confounding factors in relation to the identified evaluation
strategies? & To identify the reported factors that can distort and hence limit
the validity of the results of the SPI evaluation. To determine if
these issues are addressed and to identify possible remedies. \tabularnewline
\bottomrule
\end{tabular}%
\end{table*}

The \emph{need for this systematic review (Step 1, Figure~\ref{fig:Systematic-review-steps})
}was motivated in the introduction of this paper. In order to determine
if similar work had already been performed, we searched the Compendex,
Inspec and Google Scholar digital libraries%
\footnote{performed on 2008/11/20%
}. We used the following search string to search within keywords, title
and abstracts, using synonyms for {}``systematic review'' defined
by Biolchini et al.~\cite{de_almeida_biolchini_scientific_2007}:
\begin{verse}
((SPI \emph{OR} \textquotedbl{}Software process improvement\textquotedbl{})
\emph{AND} (\textquotedbl{}systematic review\textquotedbl{} \emph{OR}
\textquotedbl{}research review\textquotedbl{} \emph{OR} \textquotedbl{}research
synthesis\textquotedbl{} \emph{OR} \textquotedbl{}research integration\textquotedbl{}
\emph{OR} \textquotedbl{}systematic overview\textquotedbl{} \emph{OR}
\textquotedbl{}systematic research synthesis\textquotedbl{} \emph{OR}
\textquotedbl{}integrative research review\textquotedbl{} \emph{OR}
\textquotedbl{}integrative review\textquotedbl{}))
\end{verse}
None of the retrieved publications (see~\cite{unterkalmsteiner_extended_2010})
were related to our objectives which are expressed in the \emph{research
questions (Step 2)}. The research questions (Table~\ref{tab:Research-questions})
define what should be extracted from the selected publications (see
Section~\ref{sub:Data-extraction}). For example, RQ1 pertains to
how the success (or failure) of SPI initiatives is evaluated, that
is, to methods which show the impact of the initiative. Note that
with {}``evaluation strategy'' we do not refer to SPI appraisals,
such as CBA-IPI~\cite{dunaway_cmm-based_2001}, SCE~\cite{byrnes_software_1996}
or SCAMPI~\cite{_appraisal_2006}, where the organizations maturity
is assessed by its conformity to a certain industrial standard~\cite{goldenson_empirical_1996}.
We rather aim to identify the evaluation strategies which are used
to effectively show the impact of a process change. 

RQ3 investigates the measurement perspectives from which SPI initiatives
are evaluated. The perspectives (project, product and organization)
are an abstraction based on the identified metrics from RQ2. Finally,
RQ4 aims to elicit factors which may impede an accurate evaluation
of the initiative.

The aim of the \emph{review protocol (Step 3)} is to reduce potential
researcher bias and to permit a replication of the review in the future~\cite{kitchenham_guidelines_2007}.
The \emph{protocol was evaluated (Step 4)} by an independent researcher
with experience in conducting systematic reviews. According to his
feedback and our own gathered experiences during the process, we iteratively
improved the design of the review. A summary of the final protocol
is given in Sections~\ref{sub:Search-strategy} to~\ref{sub:Study-quality-assessment}.

\subsection{Search strategy\label{sub:Search-strategy}}

We followed the process depicted in Figure~\ref{fig:Search-strategy}
for the identification of papers. Figure~\ref{fig:Primary-studies-selection}
shows the selected databases and the respective number of publications
that we retrieved from each.

\begin{center}
\begin{figure}
\begin{centering}
\includegraphics[width=12pc]{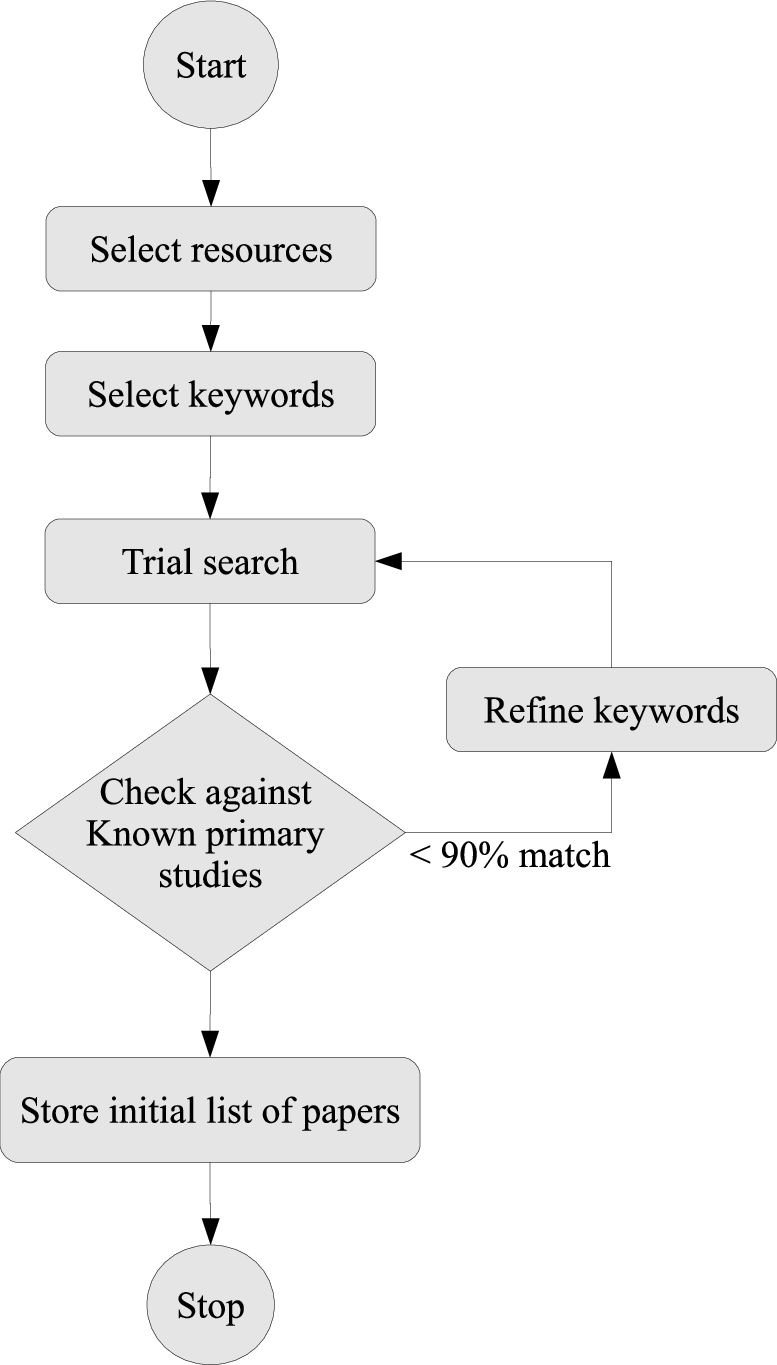}
\par\end{centering}

\caption{Search strategy\label{fig:Search-strategy}}
\end{figure}

\par\end{center}

From our research questions we derived the keywords for the search.
The search string is composed by the terms representing the population
AND intervention (Table~\ref{tab:Search-keywords}). 

\begin{table}
\caption{Search keywords\label{tab:Search-keywords}}

\centering{}\begin{tabular}{>{\raggedright}p{1.4in}>{\raggedright}p{1.8in}}
\toprule 
\multicolumn{1}{c}{Population} & \multicolumn{1}{c}{Intervention}\tabularnewline
\midrule
\midrule 
{}``process improvement'' OR “process enhancement” OR “process innovation”OR
SPI & measur{*} OR metric{*} OR success{*} OR evaluat{*} OR assess{*} OR
roi OR investment{*} OR value{*} OR cost{*} OR effect{*} OR goal{*}
OR result{*} \tabularnewline
\bottomrule
\end{tabular}%
\end{table}

In order to verify the quality of the search string, we conducted
a trial search on Inspec and Compendex. We manually identified relevant
publications from the journal {}``Software Process: Improvement and
Practice'' (SPIP) and compared them with the result-set of the trial
search. The search string captured 24 out of 31 reference publications.
Three papers were not in the result-set because Inspec and Compendex
did not index at the time of the search%
\footnote{performed on 2008/11/28%
} issues of SPIP prior to 1998. In order to capture the remaining four
publications we added the term {}``result{*}'' to the search string.

Due to the high number of publications we had to handle (10817, see
Figure~\ref{fig:Primary-studies-selection}) we decided to use a
reference management system. We opted for Zotero%
\footnote{http://www.zotero.org%
}, mainly due to its integrated capability to share and synchronize
references.

\subsection{Study selection criteria}

The main criterion for inclusion as primary study is the presentation
of empirical data showing how the discussed SPI initiative is assessed
and therefore answering the research questions (Table~\ref{tab:Research-questions}).
Both studies conducted in industry and in an academic environment
are included. Since the focus of this review is the measurement and
evaluation of SPI (see our research questions in Table~\ref{tab:Research-questions}),
general discussions on improvement approaches and comparisons of frameworks
or models were excluded. For the same reason, descriptions of practices
or tools without empirical evaluation of their application were also
not considered. Furthermore we excluded reports of {}``lessons learned''
and anecdotal evidence of SPI benefits, books, presentations, posters
and non-English texts.

\subsection{Study selection procedure\label{sub:Study-selection-procedure}}

\begin{figure}
\begin{centering}
\includegraphics[width=21pc]{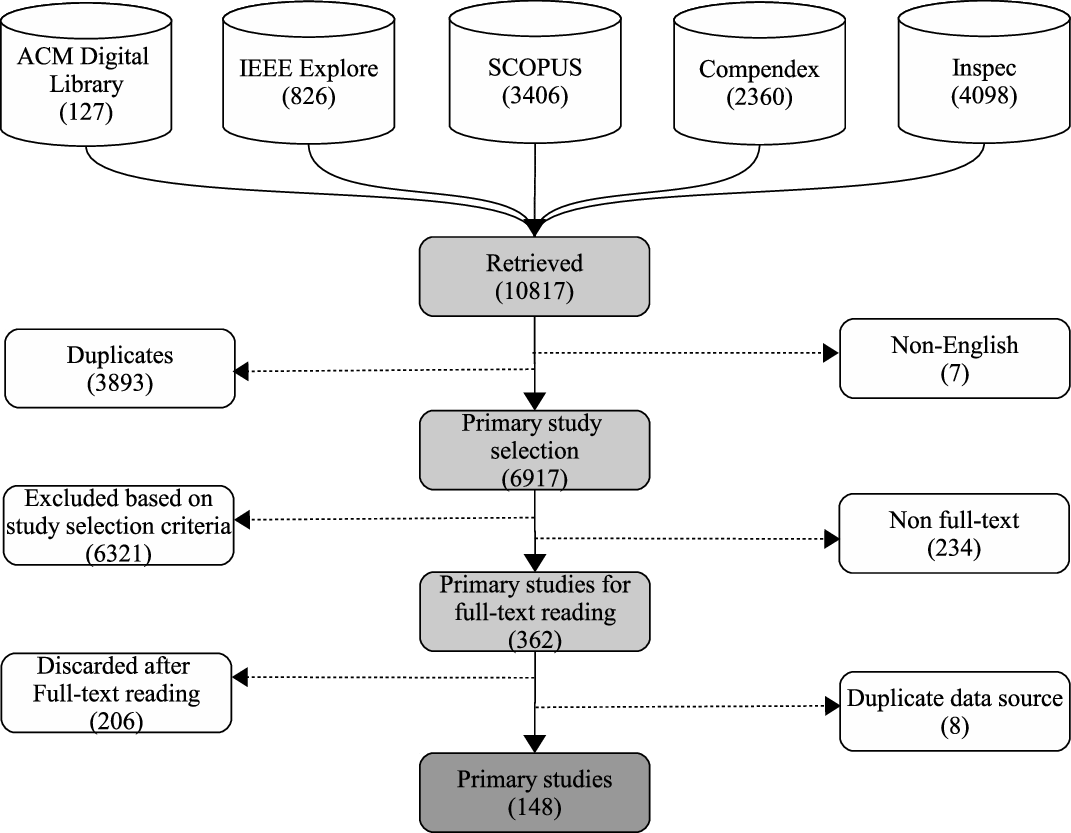}
\par\end{centering}

\caption{Primary studies selection\label{fig:Primary-studies-selection}}
\end{figure}

The systematic review \emph{procedure was first piloted (Step 5)}
in order to establish a homogeneous interpretation of the selection
criteria among the four researchers which conducted the review. The
selection criteria were applied on the title and abstract, and if
necessary, on the introduction and conclusion of the publication.
For the pilot, we assessed individually 50 randomly selected publications
from a search conducted in Inspec and Compendex. The Fleiss' Kappa~\cite{fleiss_measuring_1971}
value showed a very low agreement (0.2) among the review team. We
conducted a post-mortem analysis to unveil the causes for the poor
result. As a main reason we identified the imprecise definition of
the selection criteria and research questions, on which the decision
for inclusion was mainly based on. After a refinement of these definitions,
we conducted a second pilot on 30 randomly selected publications from
a search in SCOPUS. Furthermore, we introduced an {}``Unsure'' category
to classify publications that should be assessed by all researchers
until a consensus was reached. Fleiss' Kappa increased to a moderate
agreement (0.5), and, after that the {}``Unsure'' publications were
discussed, the inter-rater agreement improved to 0.7 (substantial
agreement according to Landis and Koch~\cite{landis_measurement_1977}),
which we considered as an acceptable level to start the selection
procedure. Figure~\ref{fig:Primary-studies-selection} illustrates
in detail how the publications retrieved from the databases were \emph{reduced
to the final primary studies (Step 6)} on which we applied the data
extraction.

As can be seen in Figure~\ref{fig:Primary-studies-selection}, from
the 10817 retrieved papers, we first discarded duplicates (by ordering
them alphabetically by their title and authors) and studies not published
in the English language. After applying the inclusion\,/\,exclusion
criteria, a total of 6321 papers were found not to be relevant and
for 234 publications we were not able to obtain a copy of the text.
This diminished the pool of papers for full-text reading to 362 papers.
In the final pool of primary studies, 148 papers remained after filtering
out studies that we found to be irrelevant after assessing the full-text
and those that reported on the same industry case studies.

\subsection{Data extraction\label{sub:Data-extraction}}

Similarly to the study selection, we distributed the workload among
four researchers. The 148 publications accepted for \emph{data extraction
(Step 7)} were randomly assigned to the extraction team (37 publications
for each member). 

We performed the data extraction in an iterative manner. Based on
the experiences reported by Staples and Niazi~\cite{staples_experiences_2007},
we expected that it would be difficult to establish \emph{a-priori}
an exhaustive set of values for all the properties. We therefore prepared
an initial extraction form with the properties listed in Table~\ref{tab:Extracted-properties},
which shows also the mapping to the respective research questions
answered by the property. For properties P1, P2, P3 and P5 a list
of expected values was established, whereas properties P4, P6 and
P7 should be extracted from the studies. Before starting the second
iteration, we reviewed the compiled extraction forms in joint meetings
and consolidated the extracted data into the categorization given
in Sections~\ref{sub:Research-method}~to~\ref{sub:Confounding-factors-data-extraction}.
In a second data extraction iteration we confirmed the established
categorization and used it for data synthesis (\emph{Step 9}) and
drawing conclusions (\emph{Step 10}).

\begin{table}
\caption{Extracted properties\label{tab:Extracted-properties}}

\centering{}\begin{tabular}{c>{\centering}p{1.5in}>{\centering}p{1.3in}}
\toprule 
ID & Property & Research question(s)\tabularnewline
\midrule
\midrule 
P1 & Research method & Overview of the studies\tabularnewline
\midrule 
P2 & Context & Overview of the studies\tabularnewline
\midrule 
P3 & SPI initiative & RQ1, RQ2, RQ3\tabularnewline
\midrule 
P4 & Success indicator and metric & RQ2\tabularnewline
\midrule 
P5 & Measurement perspective & RQ3\tabularnewline
\midrule 
P6 & Evaluation strategy & RQ1, RQ4\tabularnewline
\midrule 
P7 & Confounding factors & RQ4\tabularnewline
\bottomrule
\end{tabular}%
\end{table}

\subsubsection{Research method (P1)\label{sub:Research-method}}

We categorized the studies according to the applied research method.
Our initial strategy for the categorization was simple and straightforward:
extract the mentioned research method without interpreting the content
of the study. However, we discovered two issues with this approach.
First, the mentioned research methods were inconsistent, i.\,e.\ one
study fulfilled our understanding of a “Case study” while another
did not. Second, the research method was not mentioned at all in the
paper.

Therefore, we defined the following categories and criteria to classify
the studies consistently:
\begin{itemize}
\item \textbf{Case study }if one of the following criteria applies:

\begin{enumerate}
\item The study declares one or more research questions which are answered
(completely or partially) by applying a case study~\cite{easterbrook_selecting_2008,zelkowitz_experimental_1997}.
\item The study empirically evaluates a theoretical concept by applying
it in a case study (without necessarily explicitly stating research
questions, but having a clearly defined goal~\cite{zelkowitz_experimental_1997}).
\end{enumerate}
\item \textbf{Industry report} if the focus of the study is directed towards
reporting industrial experiences without stating research questions
or a theoretical concept which is then evaluated empirically. Usually
these studies do not mention any research method explicitly. Therefore,
instead of creating a category “N/A” (research method is not applicable),
we added this category as it complies with the {}``Project monitoring''
method described by Zelkowitz and Wallace~\cite{zelkowitz_experimental_1997}.
\item \textbf{Experiment} if the study conducts an experiment~\cite{easterbrook_selecting_2008,zelkowitz_experimental_1997}
and clearly defines its design. 
\item \textbf{Survey }if the study collects quantitative and/or qualitative
data by means of a questionnaire or interviews~\cite{easterbrook_selecting_2008,pfleeger_principles_2001,seaman_qualitative_1999}.
\item \textbf{Action research} if the study states this research method
explicitly~\cite{easterbrook_selecting_2008,davison_principles_2004}.
\item \textbf{Not stated} if the study does not define the applied research
method and it can not be derived or interpreted from reading the paper.
\end{itemize}

\subsubsection{Context (P2) \label{sub:Context}}

We categorized the studies into industry and non-industry cases. The
industry category contains studies in which the research was performed
in collaboration or embedded within industry. The non-industry category
is comprised of studies which were performed in an academic setting
or for which the research environment was not properly described.

For industrial studies we extracted the company size following the
European Recommendation 2003/361/EC~\cite{_enterprise_2009}, the
customer type (internal or external to the company) of the developed
product, the product type (pure software or embedded), and the application
domain. Furthermore, the number of projects in which the SPI initiative
was implemented and the staff-size was recorded.

Based on this information, we assessed the study quality from the
perspective of the presented research context (see QA4 in Table~\ref{tab:Quality-assessment}
in Section~\ref{sub:Study-quality-assessment}).

\subsubsection{SPI initiative (P3)\label{sub:SPI-initiative}}

We categorized the studies according to the presented SPI initiative
as follows:
\begin{itemize}
\item \textbf{Framework:} this group contains frameworks/models like CMM,
international standards like ISO/IEC~15504 (SPICE) and business management
strategies like Six Sigma. For the analysis, we further broke down
this category into:

\begin{itemize}
\item Established frameworks - CMM, CMMI, ISO/IEC~15504 (SPICE), Six-Sigma,
PSP, TSP, QIP, TQM, IDEAL, PDCA.
\item Combined frameworks - two or more established frameworks are used
in combination to implement the SPI initiative.
\item Derived frameworks - an established framework is extended or refined
to fulfill the specific needs.
\item Own framework - the study proposes a new framework without reference
to one of the established frameworks.
\item Limited framework - the framework targets only a specific process
area.
\end{itemize}
\item \textbf{Practices: }software engineering practices which can be applied
in one or more phases of the software development life-cycle (e.g.
inspections, test-driven development, etc.).
\item \textbf{Tools:} software applications that support software engineering
practices.
\end{itemize}

\subsubsection{Success indicator and metric (P4) \label{sub:Success-indicator-and-metric}}

From the inspected studies we extracted the metrics which were used
to measure the described SPI initiative. In order to get an overview
of what is actually measured, the metrics were categorized according
to “success indicators”. We did not define the classification scheme
a-priori but it emerged and evolved during the data extraction (it
was stabilized after the first iteration of the data extraction).

We use the term “success indicator” in order to describe the improvement
context in which the measurement takes place. Therefore, a “success
indicator” is an attribute of an entity (e.\,g.\ process, product,
organization) which can be used to evaluate the improvement of that
entity. The categories of success indicators is shown in Section~\ref{sub:Reported-Metrics}
(Table~\ref{tab:Success-indicators}). The identified metrics were
categorized as in the following example: (1)~The metric “Number of
defects found in peer reviews” is mapped to the “Process quality”
category as it describes the effectiveness of the peer review process
(e.\,g.\ \cite{daskalantonakis_practical_1992,russ_software_2000,ferreira_applying_2007}).
(2)~The metric “Number of defects identified after shipment\,/\,KLOC”
(e.\,g.\ \cite{daskalantonakis_practical_1992,sakamoto_analysis_1996,damian_empirical_2006})
is mapped to the “Product quality” category as the object of study
is the product itself and not the processes from which the product
originates.

The categorization of the metric is dependent on the context of the
study. The use of the metric is interpreted by understanding which
attribute is actually measured and with which intention. In some cases
this was not possible due to missing information in the description
of the metric. For example the “Defects” category contains those defect
metrics for which the given information could not be used to justify
a classification into one of the predefined quality categories (neither
product nor process).

\subsubsection{Measurement perspective (P5) \label{sub:Measurement-perspective}}

We use the concept of {}``measurement perspective'' to define and
categorize how the improvement is being assessed. Concretely, a measurement
perspective describes the view on the improvement, i.e. which entities
are measured in order to make the change visible in either a quantitative
or qualitative manner. We derived from which measurement perspective
an initiative is evaluated by interpreting the metrics which were
described in the study and from the attributes they are supposed to
measure. We defined the following measurement perspectives, based
on the five software entity types proposed by Buglione and Abran~\cite{buglione_iceberg:_2002}
(the entity types process, project and resources were bundled under
the project perspective due to the difficulty to consistently interpret
the measures identified in the reviewed studies and to avoid mis-categorization):
\begin{itemize}
\item \textbf{Project perspective}\\
The measurement is conducted during the project where the SPI initiative
takes place. Examples of metrics that are used to measure from this
perspective are productivity during the development phase, defect
rates per development phase, etc. These measures assess the entity
types process, project and resources.
\item \textbf{Product perspective}\\
The evaluation of the SPI initiatives' impact is conducted by measuring
the effect on the delivered products. An example of a metric that
is used to measure from this perspective is the number of customer
complaints after product release.
\item \textbf{Organization perspective}\\
The measurement and evaluation of the SPI initiatives' impact is
conducted organization-wide. An example of a metric that is used to
measure from this perspective is return on investment. Other qualitative
measurements such as employee satisfaction and improved business opportunities
are also measured from this perspective.
\end{itemize}

\subsubsection{Evaluation strategy (P6) \label{sub:Evaluation-strategy}}

During the first iteration of the data extraction we discovered that
many publications do not describe or define the adopted evaluation
strategy explicitly. To solve this problem, we established a categorization
of evaluation strategies based on their common characteristics (see
Section~\ref{sub:Evaluation-strategies-RQ1}, Table~\ref{tab:Evaluation-strategies}).
The categorization grew while extracting the data from the studies
and was consolidated after the first iteration of the process. In
some cases we could not identify an evaluation strategy and the publication
was categorized as \textquotedbl{}Not Stated\textquotedbl{}.

\subsubsection{Confounding factors (P7) \label{sub:Confounding-factors-data-extraction}}

\begin{figure}
\begin{centering}
\includegraphics[width=21pc]{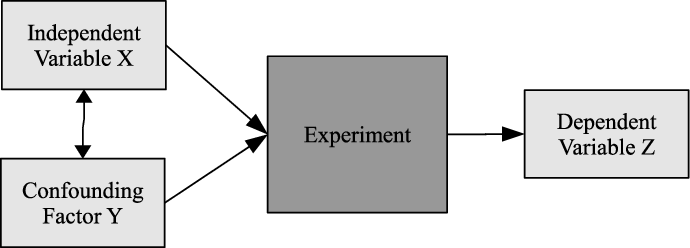}
\par\end{centering}

\caption{The influence of confounding factors\label{fig:The-influence-of-confounding-factors}}
\end{figure}

In the context of experimental design, Wohlin et al.~\cite{wohlin_empirical_2003}
defined confounding factors as “variables that may affect the dependent
variables without the knowledge of the researcher”. They represent
a threat to the internal validity of the experiment and to the causal
inferences that could be drawn since the effect of the treatment cannot
be attributed solely to the independent variable. As shown in Figure~\ref{fig:The-influence-of-confounding-factors},
both independent variables (treatments) and confounding factors represent
the input to the experiment and the assessment validity of the dependent
variables (effects) is threatened~\cite{pearl_why_1998}. 

Assuming that in the evaluation of software process improvements the
change is assessed by comparing indicators which represent an attribute
before and after the initiative has taken place, it is apparent that
the problem of confounding factors, as it is encountered in an experimental
setting, is also an issue in the evaluation of SPI initiatives. We
argue therefore that it is of paramount importance to identify potential
confounding factors in the field of software process improvement. 

Kitchenham et al.~\cite{kitchenham_case_1995} identified several
confounding factors in the context of the the evaluation of software
engineering methods and tools through case studies (see Table~\ref{tab:Confounding-factors-and-remedies}).
Similarly, we extracted from the reviewed publications any discussion
that addresses the concept of confounding factors in the context of
SPI initiatives and, if given, the chosen remedies to control the
issues.

\begin{table*}
\caption{Confounding factors and remedies (adapted from \cite{kitchenham_case_1995})
\label{tab:Confounding-factors-and-remedies}}

\centering{}\begin{tabular}{l>{\raggedright}p{3.3in}>{\raggedright}p{2.3in}}
\toprule 
\multicolumn{1}{c}{Confounding factor} & \multicolumn{1}{>{\centering}p{1.5in}}{Description} & \multicolumn{1}{c}{Remedy}\tabularnewline
\midrule
\midrule 
Learning bias & The initial effort to acquire the knowledge for using the method or
tool interferes with the evaluation of its benefits. & Separate activities aimed at learning the new technology from those
aimed at evaluating it.\tabularnewline
\midrule 
Participant bias & Attitude of the case study participants towards the new technology
(enthusiasm versus skepticism). & Select participants according to a standard staff-allocation method.\tabularnewline
\midrule 
Project bias & The projects on which the technology is evaluated differ on application
domains, e.g. embedded-systems and information systems.  & Select projects within the same application domain.\tabularnewline
\bottomrule
\end{tabular}%
\end{table*}

\subsection{Study quality assessment \label{sub:Study-quality-assessment}}

The \emph{study quality assessment (Step 8)} can be used to guide
the interpretation of the synthesis findings and to determine the
strength of the elaborated inferences~\cite{kitchenham_guidelines_2007}.
However, as also experienced by Staples and Niazi~\cite{staples_experiences_2007},
we found it difficult to assess to which extent the authors of the
studies were actually able to address validity threats. Indeed, the
quality assessment we have performed is a judgment of \emph{reporting}
rather than \emph{study} quality. We answered the questions given
in Table~\ref{tab:Quality-assessment} for each publication during
the data extraction process.

\begin{table*}
\caption{Quality assessment\label{tab:Quality-assessment}}

\centering{}\begin{tabular}{clccc}
\toprule 
ID & \multicolumn{1}{>{\centering}p{4.4in}}{Quality assessment question} & Yes & Partially & No\tabularnewline
\midrule
\midrule 
QA1 & Is the aim of the research sufficiently explained? & 138 (93.2\%) & 10 (6.8\%) & 0 (0.0\%)\tabularnewline
\midrule 
QA2 & Is the presented idea/approach clearly explained? & 115 (77.7\%) & 30 (20.3\%) & 3 (2.0\%)\tabularnewline
\midrule 
QA3 & Are threats to validity taken into consideration? & 16 (10.8\%) & 19 (12.8\%) & 113 (76.4\%)\tabularnewline
\midrule 
QA4 & Is it clear in which context the research was carried out? & 73 (49.3\%) & 54 (36.5\%) & 21 (14.2\%)\tabularnewline
\midrule 
QA5 & Are the findings of the research clearly stated? & 117 (79.0\%) & 30 (20.3\%) & 1 (0.7\%)\tabularnewline
\bottomrule
\end{tabular}%
\end{table*}

With QA1 we assessed if the authors of the study clearly state the
aims and objectives of the carried out research. This question could
be answered positively for all of the reviewed publications. With
QA2 we asked if the study provides enough information (either directly
or by referencing to the relevant literature) to give the presented
research the appropriate context and background. For almost all publications
(98\%) this could be answered positively. QA3 was checked with {}``Yes''
if validity threats were explicitly discussed, adopting the categorization
proposed by Wohlin et al.~\cite{wohlin_experimentation_2000}. The
discussion on validity threats of an empirical study increases its
credibility~\cite{perry_empirical_2000}. A conscious reflection
on potential threats and an explicit reporting of validity threats
from the researcher increases the trustworthiness of and the confidence
in the reported results. Therefore, if the study just mentioned validity
threats without properly explaining how they are identified or addressed,
the question was answered with {}``Partially''. The result of QA3
confirms the observation in~\cite{sjoberg_future_2007} that in empirical
studies the scope of validity is scarcely discussed. QA4 was answered
with {}``Yes'' if we could compile the data in the context property
of the data extraction form to a major degree (see Section~\ref{sub:Context}).
As it was pointed out by Petersen and Wohlin~\cite{petersen_context_2009},
context has a large impact on the conclusions that are drawn from
the evidence in industrial studies. However, 51.7\% of the reviewed
studies did not, or only partially, describe the context of the research.
With QA5 we assessed if the outcome of the research was properly documented.
As with QA1, this questions could be answered positively for all (except
one) study.

\subsection{Validity Threats}

We \emph{identified three potential threats to the validity (Step
11)} of the systematic review and its results.

\subsubsection{Publication bias}

Publication bias refers to the general problem that positive research
outcomes are more likely to be published than negative ones~\cite{kitchenham_guidelines_2007}.
We regard this threat as moderate, since the research questions in
this review are not geared towards the performance of a specific software
process improvement initiative for the purpose of a comparison. The
same reasoning applies to the threat of sponsoring in which certain
methods are promoted by influential organizations~\cite{kitchenham_guidelines_2007},
and negative research outcomes regarding this method are not published.
We did not restrict the sources of information to a certain publisher,
journal or conference such that it can be assumed that the breadth
of the field is covered sufficiently. However, we had to consider
the trade-off of considering as much literature as possible and, at
the same time, accumulating reliable information. Therefore we decided
not to include grey literature (technical reports, work in progress,
unpublished or not peer-reviewed publications)~\cite{kitchenham_guidelines_2007}.

\subsubsection{Threats to the identification of primary studies}

The strategy to construct the search string aimed to retrieve as many
documents as possible related to measurement and evaluation of software
process improvements. Therefore, the main metric to decide about the
quality of the search string should be the recall of the search result.
Recall is expressed as the ratio of the retrieved relevant items and
all existing relevant items~\cite{saracevic_evaluation_1995}. Since
it is impossible to know all existing relevant items, the recall of
the search string was estimated by conducting a pilot search as described
in Section~\ref{sub:Search-strategy}. This showed showed an initial
recall of 88\%, and after a refinement of the search string, a recall
of 100\%. Although the search string was exercised on a journal (SPIP)
of high relevance for this systematic review, the threat of missing
relevant articles still exists. Inconsistent terminology, in particular
in software measurement research~\cite{garcia_towards_2006}, or
use of different terminology with respect to the exercised search
string (see Table~\ref{tab:Search-keywords}) may have biased the
identification of primary studies. 

Precision, on the other hand, expresses how good the search identifies
only relevant items. Precision is defined as the ratio of retrieved
relevant items and all retrieved items~\cite{saracevic_evaluation_1995}.
We did not attempt to optimize the search string for precision. This
is clearly reflected by the final, very low, precision of 2.2\% (considering
6683 documents after the removal of duplicates and 148 selected primary
studies). This is however an expected result since recall and precision
are adversary goals, i.\,e.\ the optimization to retrieve more relevant
items (increase recall) implies usually a retrieval of more irrelevant
items too (decrease precision)~\cite{raghavan_critical_1989}. The
low precision itself represents a moderate threat to the validity
of the systematic review since it induced a considerably higher effort
in selecting the final primary studies. We addressed this threat as
explained in Section~\ref{sub:Threats-to-selection-and-data-extraction}. 

We followed two additional strategies in order to further decrease
the probability of missing relevant papers. First, during the testing
of the search string (see Section~\ref{sub:Search-strategy}), we
discovered that the bibliographic databases (Inspec and Compendex)
did not index studies published in \textquotedbl{}Software Process:
Improvement and Practice\textquotedbl{} prior to 1998. Therefore we
decided to include a third bibliographic database (SCOPUS) and also
individual publishers in the data sources (IEEE Explore and ACM Digital
Library). This led to a high number of duplicates (3893) which we
could however reliably identify by sorting the documents alphabetically
by their title and authors. Secondly, the systematic review design
was assessed for completeness and soundness by an independent researcher
with experience in conducting systematic literature reviews. 

We could not retrieve the full-text for 234 studies within the scheduled
time-frame for the systematic review. This however represents a minor
threat since this set, recalling the retrieval precision of 2.2\%,
would have contained approximately only five relevant studies.

\subsubsection{Threats to selection and data extraction consistency\label{sub:Threats-to-selection-and-data-extraction}}

Due to the scope of the systematic review, we had to develop efficient
(in terms of execution time) and effective (in terms of selection
and data extraction consistency) strategies. One of the main aims
of defining a review protocol is to reduce researcher bias~\cite{kitchenham_guidelines_2007}
by defining explicit inclusion/exclusion criteria and a data extraction
strategy. A well defined protocol increases the consistency in selection
of primary studies and in the following data extraction if the review
is conducted by multiple researchers. One approach to further increase
the validity of the review results is to conduct selection and data
extraction in parallel by several researchers and cross-check the
outcome after each phase. In the case of disagreements they should
be discussed until a final decision is achieved. Due to the large
amount of initially identified studies (10817) we found this strategy
impossible to implement within the given time-frame. Therefore, as
proposed by Brereton et al.~\cite{brereton_lessons_2007} and illustrated
in Section~\ref{sub:Study-selection-procedure}~and~\ref{sub:Data-extraction},
we piloted the paper selection and data extraction and improved the
consensus iteratively. By piloting we addressed two issues: first,
the selection criteria and the data extraction form were tested for
appropriateness, e.\,g.\ are the inclusion\,/\,exclusion criteria
too restrictive or liberal, should fields be added or removed, are
the provided options in the fields exhaustive? Second, the agreement
between the researchers could be assessed and discrepancies streamlined,
e.\,g.\ by increasing the precision of the definitions of terms.
Although it can be argued that this strategy is weaker in terms of
consistency than the previously mentioned cross-checking approach,
it was a necessary trade-off in order to fulfill the schedule and
the targeted breadth of the systematic review. 

In order to assess data extraction consistency, we performed a second
extraction on a randomly selected sample of the included primary studies.
Each researcher extracted data from 15 papers, which is slightly more
than 10\% of the total number of included studies and approximately
50\% of the studies each researcher was assigned in the first extraction.

\begin{table}
\caption{Inter- and Intra-rater agreement\label{tab:Inter--and-Intra-rater}}

\centering{}\begin{tabular}{lcc}
\toprule 
\multicolumn{1}{c}{Property} & \multicolumn{1}{c}{Inter-rater} & \multicolumn{1}{c}{Intra-rater}\tabularnewline
\midrule
\midrule 
Research method (P1) & 0.56 & 0.78\tabularnewline
\midrule 
Context (P2) & 0.90 & 0.90\tabularnewline
\midrule 
SPI initiative (P3) & 0.83 & 0.91\tabularnewline
\midrule 
Success indicator (P4) & 0.54 & 0.58\tabularnewline
\midrule 
Measurement perspective (P5) & 0.06 & 0.25\tabularnewline
\midrule 
Evaluation strategy (P6) & 0.77 & 0.47\tabularnewline
\midrule 
Confounding factors (P7) & -0.05 & -0.1\tabularnewline
\bottomrule
\end{tabular}%
\end{table}

Table~\ref{tab:Inter--and-Intra-rater} shows the Fleiss' Kappa~\cite{fleiss_measuring_1971}
value of each property that was extracted from the primary studies.
The inter-rater agreement denotes thereby the data extraction consistency
between the researchers. The intra-rater agreement gives an indication
of the repeatability of the process (the second extraction was performed
eighteen months after the original one).

Landis and Koch~\cite{landis_measurement_1977} propose the following
interpretation for Fleiss' Kappa: Almost excellent (1.0\ -\ 0.81),
Substantial (0.80\ -\ 0.61), Moderate (0.60\ -\ 0.41), Fair (0.40\ -\ 0.21),
Slight (0.20\ -\ 0), and Poor (<\ 0).

The analysis shown in Table~\ref{tab:Inter--and-Intra-rater} indicates
that in properties P5 and P7 we achieved only slight respectively
poor agreement in the data extraction validation. A potential reason
for this result on property P7 may be that confounding factors are
not explicitly mentioned in the selected primary studies and therefore
difficult to identify. In rare cases, confounding factors are mentioned
in the validity threats of the study (e.\,g.\ \cite{damian_empirical_2006})
or, more frequently, in the results discussion (e.\,g.\ \cite{paulish_case_1994,diaz_how_1997}).
A consistent extraction of property P7 is therefore rather challenging
and may be biased. 

We agreed however on the identified confounding factors (P7) and the
measurement perspective (P5) categorization, as after the original
data extraction, all involved researchers jointly discussed the results
until a consensus was reached. Hence we are confident that the reported
results in Section~\ref{sub:What-measurement-perspectives} and \ref{sub:Confounding-factors}
are internally consistent.

\section{Results and Analysis\label{sec:Results-and-Analysis}}

A total of 148 studies discuss the measurement and evaluation of SPI
initiatives. Prior to presenting the results and analysis for each
research question we give a short overview of the general characteristics
of the studies.

\subsection{Overview of the studies\label{sub:Overview-of-the-studies}}

\subsubsection{Publication year}

The reviewed papers were published between 1991 and 2008. A first
increased interest in evaluating SPI initiatives appears in the period
between 1998 and 2000 (35, 24\%). A second spike can be observed between
2005 and 2008 (55, 37\%). This seems to indicate an increased interest
in SPI and success measurement, pointing to the relevance of the area.
In addition, as a substantial part of the publications fall within
a period of four years before this review was conducted (2008), it
increases the likelihood for the results of the studies being relevant,
elevating the potential value obtained in this systematic review.

\subsubsection{Research method}

The inspected publications were classified according to the applied
research methods as defined in Section~\ref{sub:Research-method}.
Case studies (66, 45\%) and industry reports (53, 36\%) constitute
a clear majority of the studies, followed by experiments (8, 5\%),
surveys (7, 4\%), action research (1, 1\%) and a combination of action
research and experiment (1, 1\%). Also interesting to observe is that
the lack of an adequate description of the applied research methodology
prevented a categorization (12, 8\%).

\subsubsection{Study context \label{sub:Study-context}}

The study settings were categorized in industry and non-industry cases
(see Section~\ref{sub:Context}). The majority of the papers (126,
85\%) are situated in the industry category, indicating that the results
obtained from this review are based on realistic settings.

Remarkably about 50\% of the industry studies do not provide any information
on the size of the organization where the research was carried out.
The fact that considerable research effort exists to explore how to
introduce software process improvement into small and medium sized
companies~\cite{santos_implementing_2007,_improving_2006}, suggests
that company size and the available resources should be taken into
account when choosing and embarking on an SPI initiative. Omitting
that information therefore debilitates the judgment if such an initiative
is feasible in a different setting~\cite{petersen_context_2009}.
In those studies which reported the organizations size, large (>~250
employees) organizations dominate (34, 27\%) over medium (13, 10\%)
or small (<~50 employees) organizations (13, 10\%). Many publications
only provide the name of the company but they seldom provide its size
in terms of the number of employees. For well-known organizations,
this could be due to that the authors consider this information as
obvious. Another reason could be that the information was not considered
as important to report. Furthermore, confidentiality concerns are
not a valid argument for omitting context information since it is
possible to anonymize the published data~\cite{runeson_guidelines_2009}.
Indeed there are several reasons why context information such as size,
not only of the organization, but also of the unit under study can
be considered as crucial. Consider for example “A practical view of
software measurement and implementation experiences within Motorola''~\cite{daskalantonakis_practical_1992}.
The paper does not mention the size of the company. Since Motorola
is a well-known company, it is possible to get the information about
Motorola's size (at the end of 2008 it had 64000 employees~\cite{_2008_2009}).
Even if the organizations’ size at the publication date of the study
(1992) would be known, it is still difficult to judge the scope of
SPI implementation since the paper does not specify the size of, nor
in which business units the SPI initiative was implemented.

In order to improve context documentation, future SPI research should
consider to adopt the guidelines developed by Petersen and Wohlin~\cite{petersen_context_2009}.

\subsubsection{Identified SPI initiatives}

\begin{figure}
\centering{}\includegraphics[width=21pc]{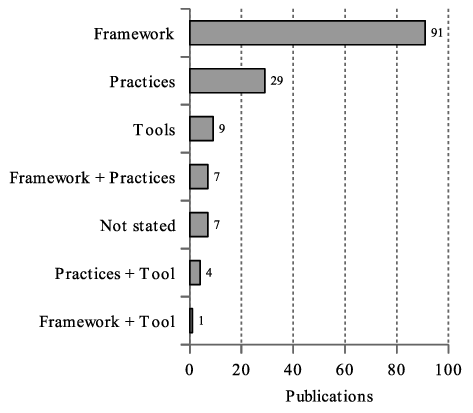}\caption{SPI initiative distribution of the publications\label{fig:SPI-initiative-distribution}}
\end{figure}

Figure~\ref{fig:SPI-initiative-distribution} shows the distribution
of the SPI initiatives according to the definition given in Section~\ref{sub:SPI-initiative}.
A detailed list of all identified initiatives can be found in the
extended material of the systematic review (see~\cite{unterkalmsteiner_extended_2010}).
Combinations of SPI initiatives (e.g. a certain practice was applied
in the context of a framework) are recorded explicitly. The “Framework”
category is predominant (91, 61\%), followed by “Practices” (29, 20\%)
and “Tools” (9, 6\%).

The scope of this systematic review is to capture any kind of process
improvement initiative and their respective approaches to evaluate
it. The holistic approach is captured by the “Framework” category
while the initiatives targeted at a limited or specific area of software
development are represented by the “Practices” and “Tools” categories.
Adding up the latter categories (i.\,e.\ the categories “Practices”,
“Tools” and “Practices~+~Tool” sum up to 42) shows that compared
to frameworks (91 studies), they are underrepresented. This suggests
that it is less common to measure and evaluate the impact of practices
and tools in the context of software process improvement research.

\begin{figure}
\centering{}\includegraphics[width=21pc]{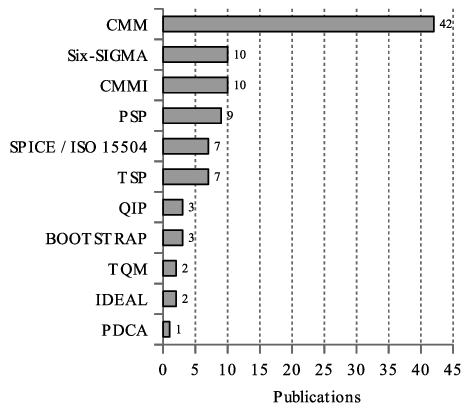}\caption{ Established framework distribution of the publications\label{fig:Established-framework-distribution}}
\end{figure}

Figure~\ref{fig:Established-framework-distribution} shows the distribution
of the established frameworks. It is of no surprise that CMM is the
most reported framework (42, 44\%) since it was introduced almost
20 years ago. The influence of the Software Engineering Institute
(SEI) can be seen here, which is also the sponsor of the CMMI, Team
and Personal Software Process (TSP/PSP) and IDEAL. SPICE (ISO/IEC~15504)
and BOOTSTRAP, software process improvement and assessment proposals
originating in Europe, are rather underrepresented. We extracted the
geographic location from the papers where the authors explicitly stated
were the study was conducted. Looking at the studies on CMM, North
America is represented 15, and Europe 9 times. On the other hand,
none of the studies on SPICE were conducted in North America. Considering
that of all identified SPI initiatives 27 studies where located in
North America and 38 in Europe, this may indicate the existence of
a locality principle, i.e. that companies adopt SPI initiatives developed
in their geographic vicinity. 

However, since the focus of the research questions is to elicit evaluation
strategies and measurements in SPI initiatives, the conclusion that
SPICE is generally less commonly used in industry cannot be drawn
from the picture; it rather means that the evaluation strategies and
measurements used in SPICE are less frequently reported by the scientific
literature.

In the following sections we answer the research questions stated
in Table~\ref{tab:Research-questions}, Section~\ref{sec:Research-Methodology}.

\subsection{Types of evaluation strategies used to evaluate SPI initiatives (RQ1)
\label{sub:Evaluation-strategies-RQ1}}

\subsubsection{Results}

The purpose of this research question was to identify the evaluation
strategies that are applied to assess the impact of an SPI initiative.
As stated in Section~\ref{sub:Evaluation-strategy}, we categorized
the strategies according to their common characteristics and established
seven categories (see Table~\ref{tab:Evaluation-strategies}). The
strategies are discussed in more detail in Section~\ref{sub:Analysis-and-Discussion-RQ1}.
The predominant evaluation strategy that we identified was {}``Pre-Post
Comparison'' (72, 49\%), followed by “Statistical Analysis” (23,
15\%). We encountered also papers where we could not identify an evaluation
strategy (21, 14\%). They were however included in the review as they
provided data points relevant to the other research questions.

\begin{table*}
\caption{Evaluation strategies\label{tab:Evaluation-strategies}}

\centering{}\begin{tabular}{>{\centering}p{1.2in}>{\raggedright}p{4.8in}c}
\toprule 
Name & \multicolumn{1}{c}{Studies} & Frequency\tabularnewline
\midrule
\midrule 
{\footnotesize Pre-Post Comparison} & \cite{regnell_market-driven_1998,visconti_measurement-based_2000,karlstraem_minimal_2005,sakamoto_analysis_1996,sommerville_empirical_2005,salo_iterative_2007,roan_pie_1999,ferreira_applying_2007,russ_software_2000,hwang_study_2005,paulish_case_1994,hollenbach_combining_1997,morad_conventional_2005,giraudo_designing_2003,ebert_e-r&d_2002,jarvinen_establishing_1999,spork_establishment_2008,von_wangenheim_experiences_2006,cuevas_impact_2007,borjesson_improve_2006,titze_improvement_2000,tanaka_improvement_1995,leung_improving_1998,anda_improving_2002,ebert_improving_2001,lane_integrating_1997,larsen_introducing_1998,dick_lessons_1997,debou_linking_2000,zettell_lipe:_2001,kautz_making_1999,winokur_measuring_1998,grable_metrics_1999,haugh_never_1991,jarvinen_integrating_1999,abrahamsson_personal_2002,cater-steel_process_2006,xiaosong_process_2008,taneike_quality_2008,kihara_quality_1998,kuntzmann-combelles_quantitative_1995,gou_quantitatively_2008,momoh_release_2006,xiaosong_research_2008,damm_results_2006,wohlwend_schlumbergers_1994,basili_sels_1995,buchman_software_1996,haley_software_1996,ahmed_results_2004,batista_spi_2000,nelson_structural_2001,lee_system_2005,ebert_technical_1999,nishiyama_technology_2000,pracchia_theav-8b_2004,ebert_impacts_2007,alagarsamy_knowledge_2007,abrahamsson_personal_2002-1,wohlin_understanding_1998,humphrey_using_1996,davis_using_2004,downey_using_2008,seshagiri_walking_2003,mcgarry_what_2002,tischer_why_2007,shah_crafting_1992,iversen_cultivation_2003,shen_measurement_2007,xu_research_2006,haessler_significant_2006} & {\footnotesize 72}\tabularnewline
\midrule 
{\footnotesize Statistical Analysis} & {\footnotesize \cite{prechelt_experiment_2001,baldassarre_industrial_2005,visaggio_assessing_2006,hollenbach_combining_1997,ramil_defining_2000,el_emam_does_1996,winkler_early_2007,nelson_evaluating_2002,suardi_how_2004,schneidewind_measuring_1999,wang_measuring_2005,seacord_measuring_2003,hayes_observe-mine-adopt_2003,he_pbr_2006,henry_quantitative_1995,hashmi_quantitative_2008,redzic_six_2006,moreau_software_2003,galinac_software_2007,ebert_understanding_2006,van_genuchten_using_2001,schalken_using_2006,freimut_industrial_2005}} & {\footnotesize 23}\tabularnewline
\midrule 
{\footnotesize Pre-Post Comparison \& Survey} & {\footnotesize \cite{damian_empirical_2006,otoya_experience:_1999,mckeown_evaluation_2000,haest_evaluation_2000,list_holistic_2005,iversen_problems_2006,damian_requirements_2005,savioja_measurement_2007,ferreira_roi_2008,johnson_software_1994}} & 10\tabularnewline
\midrule 
{\footnotesize Statistical Process Control} & {\footnotesize \cite{hollenbach_portrait_2002,french_applying_1995,daskalantonakis_practical_1992,murugappan_blending_2003,macke_optimized_2008,murugappan_quality_2000,sargut_utilization_2006,von_konsky_case_2005}} & 8\tabularnewline
\midrule 
{\footnotesize Cost-Benefit Analysis} & \cite{birk_applications_1998,trienekens_product_2001,rooijmans_software_1996,harrison_technology_1999,ebert_quest_1998} & {\footnotesize 5}\tabularnewline
\midrule 
{\footnotesize Statistical Analysis \& Survey} & {\footnotesize \cite{weiss_goal-oriented_2002,nolan_learning_1999}} & 2\tabularnewline
\midrule 
{\footnotesize Philip Crosby Associates' Approach} & {\footnotesize \cite{dion_elements_1992,dion_process_1993}} & {\footnotesize 2}\tabularnewline
\midrule 
{\footnotesize Pre-Post Comparison \& Cost-Benefit Analysis} & {\footnotesize \cite{diaz_how_1997,taek_lee_cost_2008}} & 2\tabularnewline
\midrule 
{\footnotesize Survey} & \cite{scott_practical_2001} & 1\tabularnewline
\midrule 
{\footnotesize Software Productivity Analysis Method} & \cite{bruckhaus_impact_1996} & {\footnotesize 1}\tabularnewline
\midrule 
{\footnotesize Cost-Benefit Analysis \& Survey} & {\footnotesize \cite{van_solingen_measuring_2004}} & {\footnotesize 1}\tabularnewline
\midrule 
{\footnotesize Not stated} & {\footnotesize \cite{escala_metric_1998,miller_sei_2007,canfora_applying_2006,herbsleb_systematic_1996,mcgarry_attaining_2002,lazic_cost_2008,andrade_creation_1999,ajila_empirical_2007,achatz_industrial_2003,golubic_influence_2005,hyde_intangible_2004,krasner_lessons_1996,pfleeger_maturity_1995,mcgarry_measuring_1998,valett_practical_1997,gorschek_requirements_2008,calio_software_2000,kuilboer_software_1999,biffl_software_2000,trienekens_targets_2007,el_emam_validating_2000}} & {\footnotesize 21}\tabularnewline
\bottomrule
\end{tabular}%
\end{table*}

\subsubsection{Analysis and Discussion \label{sub:Analysis-and-Discussion-RQ1}}

{}``Pre-Post Comparison'' is the most common evaluation strategy.
However, the validity of this strategy, in terms of whether the assessed
results are in causal relationship with the SPI initiative, is rarely
discussed (see Section~\ref{sub:Confounding-factors-data-extraction}
for a more detailed discussion). 

Most of the identified evaluation strategies are not specifically
designed for evaluating the outcome of SPI initiatives. However, an
exception is given by the Philip Crosby Associates’ Approach, which
suggests explicitly what to evaluate~\cite{crosby_quality_1984}.
The majority of the found evaluation strategies are very generic in
nature and different organizations applied those methods for measuring
different success indicators based on the organizational needs and
contexts. This indicates that there is a shortcoming in the used methods
to evaluate the outcome of SPI initiative in a consistent and appropriate
way, and supports the demand~\cite{iversen_problems_2006} for a
comprehensive measurement framework for SPI.

\textbf{Pre-Post Comparison:} The outcome of SPI initiatives is evaluated
by comparing the success indicators’ values before and after the SPI
initiatives took place. Hence, for the {}``Pre-Post Comparison''
of success indicators it is necessary to setup a baseline from which
the improvements can be measured~\cite{rozum_concepts_1993}. The
major difficulty here is to identify reasonable baseline values. One
strategy could be to use the values from a very successful project
or product (either internal or external to the organization) and benchmark
the improvement against those. Accordingly, the baseline would represent
the target that is aimed for in the improvement. Benchmarking in this
way is useful if no historical data of successful projects or products
is available. However, the performance of the improvement initiative
cannot be deduced by comparing against a target baseline since the
previous status is unknown and therefore the target may merely serve
as an indication. Therefore, for evaluating the effect of improvement
initiatives, historical data against which the actual performance
can be compared is essential. An example that illustrates how a baseline
for organizational performance can be constructed is given by Paulish
and Carleton~\cite{paulish_case_1994}. Organizations with an established
measurement program will have less difficulty to establish a baseline
than organizations with a newly instantiated or even not yet started
program~\cite{paulish_case_1994}. 

Baselines are also essential in statistical process control (SPC)
where the variation of a specific process attribute relative to a
baseline is interpreted as instability and therefore a possible cause
of quality issues of the resulting product. Hollenbach and Smith~\cite{hollenbach_portrait_2002}
exemplify the establishment of baselines for SPC. Furthermore, the
statistical techniques presented by Henry et al.~\cite{henry_quantitative_1995}
can be used to create baselines of quality and productivity measurements.

\textbf{Statistical Analysis and Statistical Process Control (SPC):}
Statistical analysis includes descriptive statistics where data are
summarized numerically (e.\,g.\ mean, median, mode) or graphically
(e.\,g.\ charts and graphs). Statistical analysis can also be done
by inferential statistics by drawing inferences about the larger population
through hypothesis testing, estimates of numerical characteristics
(estimation), descriptions of association (correlation), or modeling
of relationships (regression). One application of statistical techniques
is to strengthen the validity of the collected measurements~\cite{rocha_strengthening_????}.
Another common application is found in SPC which aim is to measure
and analyze the variation in processes. Time series analysis, as promoted
by SPC, can provide information when an improvement should be carried
out and determine the efficacy of the process changes~\cite{caivano_continuous_2005}. 

As proposed by Henry et al.~\cite{henry_quantitative_1995}, several
statistical techniques can be applied to evaluate the effectiveness
of software process improvement in terms of increased estimation accuracy,
product quality and customer satisfaction. The described methods are
multiple regression, rank correlation and chi-square tests of independence
in two-way contingency tables, which, when applied repeatedly over
time can show the effectiveness of process improvements statistically~\cite{henry_quantitative_1995}.
However, care must be taken when applying these techniques since a
single method alone may not show the true impact of the initiative
and wrong conclusions could be drawn~\cite{henry_quantitative_1995}.
Furthermore Henry et al.~\cite{henry_quantitative_1995} objected
that in some cases the process improvement must be very effective
in order to show significant alterations in the statistical evaluation
results. Statistical methods are also used to assess process stability
which is regarded as an important aspect of organizational capability~\cite{schneidewind_measuring_1999}.
In order to evaluate stability, the authors propose trend, change
and shape metrics which can be used in the short- and long-term and
are analyzed by visual inspection of the data summarized by descriptive
statistics (e.\,g.\ histograms and trend diagrams). 

Ramil and Lehman~\cite{ramil_defining_2000} discuss the assessment
of process improvement from the viewpoint of software evolution. The
authors propose a statistical technique to determine whether productivity
(or any other process or product attribute) changes significantly
over a long period of time. The aim of the presented CUSUM (cumulative
sum) test is to systematically explore data points which highlight
changes in the evolutionary behavior. Although this can also be done
by visual inspection of trends (as it was proposed by Schneidewind~\cite{schneidewind_measuring_1999}),
a change detection algorithm is considered as less error-prone and
is particularly useful when assessing the impact of process improvement
initiatives and when analyzing whether the performance of processes
has changed~\cite{ramil_defining_2000}. 

An interesting approach to address the issue of certain confounding
factors using statistical techniques is presented by Schalken et al.~\cite{schalken_using_2006}.
The authors illustrate how Cost-Model Comparison, based on a linear
regression equation, can account for the factor of project size when
evaluating the effect of a process improvement on productivity (the
same method is also proposed by Alagarsamy et al.~\cite{alagarsamy_knowledge_2007}).
A second issue, namely the comparison of projects from different departments
to assess productivity improvement is addressed by the Hierarchical
Model Approach. Projects originating from different departments in
an organization are not directly comparable since they are either
specialized on a group of products, a specific technology or have
employees with different skills~\cite{schalken_using_2006}. Both
the Cost-Model Comparison and the Hierarchical Model Approach can
be used to prevent erroneous conclusions about the impact of the process
improvement initiative by considering context. Unfortunately, as we
have shown in Section~\ref{sub:Study-context}, the context in which
the improvement initiatives are evaluated, is seldom presented completely.
It is therefore difficult to judge in such cases if the reported improvement
can be attributed to the initiative.

\textbf{Survey:} In the context of this work, a survey is defined
as any method to collect, compare and evaluate quantitative or qualitative
data from human subjects. A survey can be conducted by interviews
or questionnaires, targeting employees affected by the process improvement
initiative or customers of the organization. Surveys can be an effective
mean to assess the changes introduced in an improvement effort since
after all, the development of software is a human-intensive task.
The feedback provided by employees can therefore be used to improve
the understanding of the effects caused by the introduced changes
and to steer future improvements. Gathering information from customers,
on the other hand, can provide insight how the improvement affects
the quality of products or services as perceived by their respective
users. This can be valuable to assess external quality characteristics,
such as integrity, reliability, usability, correctness, efficiency
and interoperability~\cite{fitzpatrick_usable_1998}, which otherwise
would be difficult to evaluate. The analysis of the improvement participants'
feedback can be valuable if historical data for comparison is not
available or if its quality\,/\,completeness limits the evaluability
of the improvement. A systematic method to assess the effects caused
by an improvement initiative is described by Pettersson~\cite{pettersson_practitioners_2008}.
The approach can be useful if no or only limited historical data is
available to construct a baseline which can serve as a reference point
for the improvement evaluation. The post-evaluation is based on the
expert opinion of the directly involved personnel which compares the
improved process with the previous one. This lightweight process improves
the visibility on the effects of the undertaken improvement initiative
and provides also information on how the change was experienced by
the involved roles. The method could be enhanced by integrating the
concept of \textquotedbl{}contribution percentages\textquotedbl{}
as it was proposed by van Solingen~\cite{van_solingen_measuring_2004}.
The idea is to let the experts assess how much the initiative actually
contributed to the improvement, i.\,e.\ provide the possibility
to express that only a fraction of the change is attributable to the
initiative and other factors have also contributed to the enhancement.
Such an approach could also support the identification of potential
confounding factors (see Section~\ref{sub:Confounding-factors}). 

Besides by the expert opinion of employees, it is also possible to
evaluate the effects of the improvement by querying customers. Quality
of service surveys could be sent periodically to customers, illustrating
the effects of the adapted or new process from the customer perspective~\cite{mckeown_evaluation_2000}.

\textbf{Cost-Benefit Analysis:} Evaluating an improvement initiative
with a cost-benefit measure is important since the allocated budget
for the program must be justifiable in order not to risk its continuation~\cite{van_solingen_measuring_2004,kitchenham_software_1996}.
Furthermore, it is necessary to avoid loss of money and to identify
the most efficient investment opportunities~\cite{van_solingen_measuring_2004}.
When assessing cost, organizations should also consider other resources
than pure effort (which can be relatively easily measured), e.\,g.\ office
space, travel, computer infrastructure~\cite{van_solingen_measuring_2004},
training, coaching, additional metrics, additional management activities,
process maintenance~\cite{ebert_quest_1998}. Activity Based Costing
helps to relate certain activities with the actual spent effort~\cite{ebert_quest_1998}.
Since cost and effort data can be collected in projects, they must
not be estimated~\cite{ebert_quest_1998}. On the other hand, the
thereby obtained values are still an approximation and estimations
of both costs and benefits are inevitable~\cite{van_solingen_measuring_2004}.
Since it is usually enough to know the ROI's relative value (positive,
balanced or negative), perfect accuracy is not required as long as
the involved stakeholders agree on the procedure how to assess it~\cite{van_solingen_measuring_2004}.
Direct benefits and especially indirect and intangible benefits are
best assessed by multiple stakeholders~\cite{van_solingen_measuring_2004};
some of the difficult to quantify benefits are: customer satisfaction,
improved market share due to improved quality, reduced time-to-deliver
and accuracy, feature-cost reduction, opportunity costs, reduced maintenance
in follow-up projects, better reusability, employee satisfaction,
increased resource availability~\cite{ebert_quest_1998}. A useful
technique to support the estimation is the so-called \textquotedbl{}what-if-not\textquotedbl{}
analysis~\cite{van_solingen_measuring_2004}. Project managers could
be asked to estimate how much effort was saved due to the implemented
improvement in follow-up projects. The saved effort would then be
accounted as a benefit. Another strategy would be to estimate the
\textquotedbl{}worth\textquotedbl{} of a certain improvement, e.\,g.\ asking
managers how many training days would they invest to increase employee
motivation and quantify the cost of such a training program~\cite{van_solingen_measuring_2004}.

\textbf{Philip Crosby Associates' Approach:} This method is derived
from Philip Crosby's\emph{ Cost of Quality} idea~\cite{crosby_quality_1984}.
It is based on distinguishing the cost of doing it right the first
time (performance costs) from the cost of rework (non-conformance
costs). The cost of quality is determined by the sum of appraisal,
prevention and rework costs~\cite{dion_elements_1992}. The improvement
is evaluated by a reduction of rework costs over a longer period of
time (several years, as shown in~\cite{dion_elements_1992} and~\cite{dion_process_1993}).
This method is similar to Cost-Benefit Analysis but particularly tailored
to software process improvement evaluation.

\textbf{Software Productivity Analysis Method (SPAM):} SPAM provides
a way of defining productivity models and evaluation algorithms to
calculate the productivity of all possible combinations of an observed
phenomenon (process, project size, technology etc.)~\cite{bruckhaus_impact_1996,bruckhaus_quantitative_1997}.

\subsection{Reported metrics for evaluating the SPI initiatives (RQ2)\label{sub:Reported-Metrics}}

\subsubsection{Results}

The purpose of this research was to identify the used metrics and
success indicators (see Section~\ref{sub:Success-indicator-and-metric})
in SPI evaluations.

\begin{table*}
\caption{Success indicators\label{tab:Success-indicators}}

\centering{}\begin{tabular}{>{\raggedright}p{1in}>{\raggedright}p{2in}>{\raggedright}p{2.5in}>{\centering}p{0.6in}}
\toprule 
\multicolumn{1}{c}{Success indicator} & \multicolumn{1}{c}{Description} & \multicolumn{1}{c}{Studies} & Frequency\tabularnewline
\midrule
\midrule 
{\footnotesize Process Quality} & {\footnotesize The quality that indicates process performance and
is not related to the product. The metrics for this success indicator
are dependent on the type of process they measure.} & {\footnotesize \cite{escala_metric_1998,sommerville_empirical_2005,otoya_experience:_1999,baldassarre_industrial_2005,canfora_applying_2006,ferreira_applying_2007,french_applying_1995,daskalantonakis_practical_1992,russ_software_2000,visaggio_assessing_2006,murugappan_blending_2003,lazic_cost_2008,giraudo_designing_2003,lee_development_2005,winkler_early_2007,ebert_e-r&d_2002,jarvinen_establishing_1999,spork_establishment_2008,nelson_evaluating_2002,haest_evaluation_2000,von_wangenheim_experiences_2006,list_holistic_2005,diaz_how_1997,leung_improving_1998,larsen_introducing_1998,zettell_lipe:_2001,savioja_measurement_2007,schneidewind_measuring_1999,wang_measuring_2005,winokur_measuring_1998,macke_optimized_2008,he_pbr_2006,cater-steel_process_2006,trienekens_product_2001,kihara_quality_1998,murugappan_quality_2000,kuntzmann-combelles_quantitative_1995,henry_quantitative_1995,gou_quantitatively_2008,damian_requirements_2005,gorschek_requirements_2008,damm_results_2006,basili_sels_1995,redzic_six_2006,biffl_software_2000,moreau_software_2003,batista_spi_2000,trienekens_targets_2007,ebert_technical_1999,abrahamsson_personal_2002-1,wohlin_understanding_1998,van_genuchten_using_2001,sargut_utilization_2006,seshagiri_walking_2003,mcgarry_what_2002,xu_research_2006}} & {\footnotesize 57}\tabularnewline
\midrule 
{\footnotesize Estimation Accuracy} & {\footnotesize The deviation between the actual and planned values
of other attributes measurements. Examples of attributes commonly
estimated are schedule, effort, size and productivity.} & {\footnotesize \cite{von_konsky_case_2005,regnell_market-driven_1998,escala_metric_1998,sakamoto_analysis_1996,sommerville_empirical_2005,damian_empirical_2006,prechelt_experiment_2001,roan_pie_1999,hollenbach_portrait_2002,ferreira_applying_2007,french_applying_1995,daskalantonakis_practical_1992,hwang_study_2005,herbsleb_systematic_1996,mcgarry_attaining_2002,paulish_case_1994,taek_lee_cost_2008,shah_crafting_1992,iversen_cultivation_2003,jarvinen_establishing_1999,von_wangenheim_experiences_2006,cuevas_impact_2007,anda_improving_2002,achatz_industrial_2003,zettell_lipe:_2001,kautz_making_1999,wang_measuring_2005,grable_metrics_1999,abrahamsson_personal_2002,iversen_problems_2006,dion_process_1993,hashmi_quantitative_2008,kuntzmann-combelles_quantitative_1995,henry_quantitative_1995,damian_requirements_2005,gorschek_requirements_2008,ferreira_roi_2008,wohlwend_schlumbergers_1994,haley_software_1996,calio_software_2000,kuilboer_software_1999,johnson_software_1994,moreau_software_2003,rooijmans_software_1996,ahmed_results_2004,ebert_technical_1999,pracchia_theav-8b_2004,ebert_impacts_2007,abrahamsson_personal_2002-1,davis_using_2004,mcgarry_what_2002,mcgarry_measuring_1998,ebert_understanding_2006,el_emam_validating_2000,seshagiri_walking_2003,xu_research_2006}} & {\footnotesize 56}\tabularnewline
\midrule 
{\footnotesize Productivity} & {\footnotesize The performance of the development team in terms of
its efficiency in delivering the required output.} & {\footnotesize \cite{von_konsky_case_2005,sakamoto_analysis_1996,damian_empirical_2006,prechelt_experiment_2001,baldassarre_industrial_2005,miller_sei_2007,french_applying_1995,daskalantonakis_practical_1992,herbsleb_systematic_1996,mcgarry_attaining_2002,paulish_case_1994,iversen_cultivation_2003,ramil_defining_2000,lee_development_2005,ajila_empirical_2007,jarvinen_establishing_1999,cuevas_impact_2007,titze_improvement_2000,lane_integrating_1997,kautz_making_1999,pfleeger_maturity_1995,grable_metrics_1999,valett_practical_1997,iversen_problems_2006,dion_process_1993,xiaosong_process_2008,taneike_quality_2008,gou_quantitatively_2008,hashmi_quantitative_2008,damian_requirements_2005,ferreira_roi_2008,wohlwend_schlumbergers_1994,buchman_software_1996,haley_software_1996,rooijmans_software_1996,ahmed_results_2004,lee_system_2005,nishiyama_technology_2000,bruckhaus_impact_1996,alagarsamy_knowledge_2007,schalken_using_2006,mcgarry_what_2002,hwang_study_2005,shah_crafting_1992,diaz_how_1997,nolan_learning_1999,mcgarry_measuring_1998,humphrey_using_1996,el_emam_validating_2000,shen_measurement_2007,xu_research_2006,bellini_cmm_2006}} & {\footnotesize 52}\tabularnewline
\midrule 
{\footnotesize Product Quality} & {\footnotesize This list shows studies in which }{\footnotesize \par}

{\footnotesize we identified measures for product quality in general.
Studies, in which a quality attribute according to the ISO 9126-1
standard was mentioned, are shown in Table \ref{tab:Product-Quality-Success-Indicators}.} & {\footnotesize \cite{regnell_market-driven_1998,sakamoto_analysis_1996,damian_empirical_2006,baldassarre_industrial_2005,herbsleb_systematic_1996,list_holistic_2005,titze_improvement_2000,tanaka_improvement_1995,golubic_influence_2005,lane_integrating_1997,larsen_introducing_1998,dick_lessons_1997,abrahamsson_personal_2002,taneike_quality_2008,ferreira_roi_2008,wohlwend_schlumbergers_1994,haley_software_1996,lee_system_2005,ebert_technical_1999,ebert_impacts_2007,abrahamsson_personal_2002-1,otoya_experience:_1999,el_emam_does_1996,von_wangenheim_experiences_2006,nolan_learning_1999,mcgarry_measuring_1998,xiaosong_research_2008,seshagiri_walking_2003}} & {\footnotesize 47 (28 + Table \ref{tab:Product-Quality-Success-Indicators}}\tabularnewline
\midrule 
{\footnotesize Effort } & {\footnotesize The effort of the development team in developing the
product.} & {\footnotesize \cite{escala_metric_1998,sakamoto_analysis_1996,sommerville_empirical_2005,baldassarre_industrial_2005,salo_iterative_2007,miller_sei_2007,visaggio_assessing_2006,paulish_case_1994,giraudo_designing_2003,winkler_early_2007,titze_improvement_2000,tanaka_improvement_1995,achatz_industrial_2003,larsen_introducing_1998,dick_lessons_1997,debou_linking_2000,kautz_making_1999,wang_measuring_2005,grable_metrics_1999,hayes_observe-mine-adopt_2003,jarvinen_integrating_1999,trienekens_product_2001,kuntzmann-combelles_quantitative_1995,gorschek_requirements_2008,basili_sels_1995,calio_software_2000,rooijmans_software_1996,trienekens_targets_2007,ebert_technical_1999,nishiyama_technology_2000,wohlin_understanding_1998,van_genuchten_using_2001,downey_using_2008,sargut_utilization_2006,mcgarry_what_2002,karlstraem_minimal_2005,hwang_study_2005,von_wangenheim_experiences_2006,suardi_how_2004,haessler_significant_2006,xu_research_2006}} & {\footnotesize 41}\tabularnewline
\midrule 
{\footnotesize Defects} & {\footnotesize This success indicator is to group metrics that are
solely intended to measure the defects without relating them to quality.} & {\footnotesize \cite{escala_metric_1998,freimut_industrial_2005,miller_sei_2007,french_applying_1995,daskalantonakis_practical_1992,mcgarry_attaining_2002,paulish_case_1994,hollenbach_combining_1997,cuevas_impact_2007,ebert_improving_2001,achatz_industrial_2003,wang_measuring_2005,grable_metrics_1999,haugh_never_1991,iversen_problems_2006,murugappan_quality_2000,gou_quantitatively_2008,redzic_six_2006,buchman_software_1996,rooijmans_software_1996,galinac_software_2007,nelson_structural_2001,ebert_technical_1999,pracchia_theav-8b_2004,van_genuchten_using_2001,davis_using_2004,downey_using_2008,sargut_utilization_2006,karlstraem_minimal_2005,haest_evaluation_2000,suardi_how_2004,humphrey_using_1996,tischer_why_2007,shen_measurement_2007,xu_research_2006}} & {\footnotesize 35}\tabularnewline
\bottomrule
\end{tabular}%
\end{table*}

\begin{table*}
\caption{Success Indicators (Continued)\label{tab:Success-Indicators-(Continued)}}

\centering{}\begin{tabular}{>{\raggedright}p{1in}>{\raggedright}p{2in}>{\raggedright}p{2.5in}>{\centering}p{0.6in}}
\toprule 
\multicolumn{1}{c}{Success indicator} & \multicolumn{1}{c}{Description} & \multicolumn{1}{c}{Studies} & \multicolumn{1}{c}{Frequency}\tabularnewline
\midrule
\midrule 
{\footnotesize Cost} & {\footnotesize The cost in terms of the resources that is required
in developing the product (monetary expenses).} & {\footnotesize \cite{miller_sei_2007,daskalantonakis_practical_1992,morad_conventional_2005,taek_lee_cost_2008,dion_elements_1992,ajila_empirical_2007,mckeown_evaluation_2000,list_holistic_2005,tanaka_improvement_1995,ebert_improving_2001,achatz_industrial_2003,golubic_influence_2005,lane_integrating_1997,dion_process_1993,murugappan_quality_2000,gorschek_requirements_2008,basili_sels_1995,buchman_software_1996,haley_software_1996,batista_spi_2000,nelson_structural_2001,bruckhaus_impact_1996,wohlin_understanding_1998,nelson_evaluating_2002,von_wangenheim_experiences_2006,winokur_measuring_1998,damm_results_2006}} & {\footnotesize 27}\tabularnewline
\midrule 
{\footnotesize Time-to-Market} & {\footnotesize The time that it takes to deliver a product to the
market from its conception time.} & {\footnotesize \cite{sommerville_empirical_2005,miller_sei_2007,mcgarry_attaining_2002,paulish_case_1994,hollenbach_combining_1997,ebert_e-r&d_2002,weiss_goal-oriented_2002,list_holistic_2005,ebert_improving_2001,achatz_industrial_2003,golubic_influence_2005,lane_integrating_1997,kautz_making_1999,jarvinen_integrating_1999,momoh_release_2006,wohlwend_schlumbergers_1994,buchman_software_1996,trienekens_targets_2007,ebert_technical_1999,ebert_impacts_2007,downey_using_2008,shah_crafting_1992,diaz_how_1997,zettell_lipe:_2001,winokur_measuring_1998,mcgarry_measuring_1998,haessler_significant_2006}} & {\footnotesize 27}\tabularnewline
\midrule 
{\footnotesize Other Qualitative Success Indicators} & {\footnotesize Examples are: staff morale, employee satisfaction,
quality awareness} & {\footnotesize \cite{damian_empirical_2006,roan_pie_1999,herbsleb_systematic_1996,mcgarry_attaining_2002,andrade_creation_1999,list_holistic_2005,borjesson_improve_2006,achatz_industrial_2003,hyde_intangible_2004,krasner_lessons_1996,wang_measuring_2005,scott_practical_2001,iversen_problems_2006,dion_process_1993,trienekens_product_2001,momoh_release_2006,calio_software_2000,rooijmans_software_1996,canfora_applying_2006,shah_crafting_1992,xiaosong_research_2008,el_emam_validating_2000,shen_measurement_2007}} & {\footnotesize 24}\tabularnewline
\midrule 
{\footnotesize Return-On-Investment} & {\footnotesize The value quantified by considering the benefit and
cost of software process improvement.} & {\footnotesize \cite{miller_sei_2007,roan_pie_1999,birk_applications_1998,paulish_case_1994,taek_lee_cost_2008,lazic_cost_2008,borjesson_improve_2006,golubic_influence_2005,larsen_introducing_1998,krasner_lessons_1996,murugappan_quality_2000,kuntzmann-combelles_quantitative_1995,momoh_release_2006,gorschek_requirements_2008,ferreira_roi_2008,nishiyama_technology_2000,ebert_quest_1998,diaz_how_1997,van_solingen_measuring_2004,damm_results_2006,shen_measurement_2007,haessler_significant_2006}} & {\footnotesize 22}\tabularnewline
\midrule 
{\footnotesize Customer Satisfaction (Qualitative)} & {\footnotesize The level of customer expectation fulfillment by the
organization's product and service. Customer satisfaction measurement
is divided into two types, qualitative and quantitative.} & {\footnotesize \cite{miller_sei_2007,herbsleb_systematic_1996,iversen_cultivation_2003,mckeown_evaluation_2000,list_holistic_2005,achatz_industrial_2003,golubic_influence_2005,savioja_measurement_2007,wang_measuring_2005,iversen_problems_2006,gou_quantitatively_2008,momoh_release_2006,ferreira_roi_2008,kuilboer_software_1999,johnson_software_1994,moreau_software_2003,shah_crafting_1992,el_emam_does_1996,nolan_learning_1999,el_emam_validating_2000}} & {\footnotesize 20}\tabularnewline
\midrule 
{\footnotesize Customer Satisfaction (Quantitative)} &  & {\footnotesize \cite{daskalantonakis_practical_1992,mckeown_evaluation_2000,debou_linking_2000,wohlwend_schlumbergers_1994,moreau_software_2003,batista_spi_2000,hwang_study_2005}} & {\footnotesize 7}\tabularnewline
\midrule 
{\footnotesize Other Quantitative Success Indicators} & {\footnotesize This success indicator is to group metrics that measure
context-specific attributes which are not part of any of the above
success indicators (e.g. employee satisfaction, innovation)} & {\footnotesize \cite{escala_metric_1998,weiss_goal-oriented_2002,list_holistic_2005,savioja_measurement_2007,harrison_technology_1999,suardi_how_2004}} & {\footnotesize 6}\tabularnewline
\midrule 
{\footnotesize Other Quality Attributes} &  & {\footnotesize \cite{taek_lee_cost_2008,iversen_cultivation_2003,seacord_measuring_2003,wohlwend_schlumbergers_1994}} & {\footnotesize 4}\tabularnewline
\bottomrule
\end{tabular}%
\end{table*}

Table~\ref{tab:Success-indicators} and Table~\ref{tab:Success-Indicators-(Continued)}
show the frequency of the identified success indicators in the inspected
studies. “Process Quality” (57, 39\%) was the most observed success
indicator, followed by “Estimation Accuracy” (56, 38\%), “Productivity”
(52, 35\%) and “Product Quality” (in total 47 papers, 32\%, considering
also those from Table~\ref{tab:Product-Quality-Success-Indicators}). 

We differentiated the {}``Product Quality'' success indicators based
on the ISO~9126-1 standard. The identified studies are shown in Table~\ref{tab:Product-Quality-Success-Indicators}.
Two points have to be noted. First, we added “Reusability”, which
is not defined as a product quality attribute by ISO~9126-1, to the
quality attributes. Furthermore, if the study did not explicitly state,
or sufficiently describe, which quality attribute is measured, we
mapped the study to the general “Product Quality” category (see Table~\ref{tab:Success-indicators}).

“Reliability” was the most observed success indicator for the product
quality characteristics, followed by “Maintainability” and “Reusability''.

Table~\ref{tab:Estimation-Accuracy-Success-Indicators} shows the
categorization of estimation accuracy indicators. The “Others” category
contains again estimation accuracy metrics which could not be mapped
to the specific categories. “Schedule” (37, 25\%) is by far the most
observed success indicator for estimation accuracy. On the other hand,
assuming that “Cost” can be expressed in terms of “Effort” and vice
versa, combining them shows that their number of observations (35,
24\%) is comparable to that one of “Schedule”. “Size” (10, 7\%), “Productivity”
and “Quality” (2 papers each, 1\%) fall behind.

We also distinguished how customer satisfaction is assessed (Table~\ref{tab:Success-Indicators-(Continued)}).
Qualitative customer satisfaction is largely assessed by questionnaires,
while quantitative customer satisfaction is recorded by objective
measures (e.g. New open problems~=~total new post-release problems
opened during the month). 

The “Other Qualitative/Quantitative Success Indicator” categories
contain indicators such as “Team morale”, “Employee motivation” or
“Innovation” which were explicitly mentioned in the studies as indicators
for improvement but could not be mapped into the classification.

\subsubsection{Analysis and Discussion \label{sub:Analysis-and-Discussion-Metrics}}

The main incentive behind the embarkment of an SPI initiative is to
increase quality and to decrease cost and schedule~\cite{raffo_role_1997,van_solingen_calculating_2006,staples_systematic_2008}.
In order to evaluate the success of such an initiative it is crucial
to assess the improvement's effects. Table~\ref{tab:Success-indicators}
and Table~\ref{tab:Success-Indicators-(Continued)} list the success
indicators we identified in this systematic review. We mapped the
improvement goals of quality, cost and schedule with these success
indicators:
\begin{itemize}
\item Quality (“Process Quality” \& “Product Quality” \& {}``Other Quality
Attributes'') was found in 92 papers, 62\% 
\item Cost (“Effort” \& “Cost”) was found in 61 papers, 41\% 
\item Schedule (“Time-to-market”) was found in 27 papers, 18\%
\end{itemize}
This shows that quality is the most measured attribute, followed by
cost and schedule. Drawing an analogy with the time-cost-performance
triangle ~\cite{atkinson_project_1999,kerzner_project_2009}, which
reflects that the three properties are interrelated and it is not
possible to optimize all three at the same time, the unbalanced number
in the identified success indicators suggests that this is also true
for what is actually measured in SPI initiatives.

Furthermore, in order to accurately calculate the financial benefits
of an SPI initiative, it is necessary to take all three attributes
into account~\cite{raffo_role_1997}. The low occurrence of “Return-on-investment”
(22, 15\%) as success indicator suggests that it is seldom used to
increase the visibility of the improvement efforts. It has been shown,
however, that {}``Return-on-investment'' can be used to communicate
the results of an SPI initiative to the various stakeholders~\cite{rico_roi_2004}
(see Section~\ref{sub:Analysis-and-Discussion-RQ1}, Cost-Benefit
Analysis for are more in-depth discussion about {}``Return-on-Investment'').

\textbf{Product Quality:} As shown in Table~\ref{tab:Product-Quality-Success-Indicators},
we categorized success indicators according to ISO~9126-1 product
quality attributes. The main incentive to analyze the success indicators
from this perspective is that those attributes may have a different
weight, depending on the stakeholder. A developer may rate “Maintainability”,
“Reusability” and “Portability” (internal quality attributes) higher
than the products customer. “Reliability”, “Usability”, “Functionality”
and “Efficiency” on the other hand are the external quality attributes
of the product which are potentially more important to the customer~\cite{bevan_quality_1999,stavrinoudis_comparing_2008}.
The measurement of internal quality attributes can be applied efficiently,
with a low error frequency and cost~\cite{stavrinoudis_comparing_2008}.
It is therefore of no surprise that internal attributes are measured
more frequently than external ones (see Table~\ref{tab:Product-Quality-Success-Indicators}).
Interestingly, “Reliability” is measured far more often as compared
to the other three external attributes. This is explained by looking
at the used measures in these studies to express {}``Reliability'',
which in the majority are based on product failures reported by the
customer and therefore relatively easy to collect and evaluate. On
the other hand, {}``Usability'' which is considered as difficult
to measure~\cite{seffah_usability_2006,bevan_usability_1994,mcnamara_functionality_2006},
is also seldom assessed in the context of process improvement (see
Table~\ref{tab:Product-Quality-Success-Indicators}).

\textbf{Customer Satisfaction:} Customer satisfaction can be used
to determine software quality~\cite{pressman_software_2001} since
it is commonly considered as an accomplishment of quality management~\cite{ho-won_measuring_2004}.
An increased product quality could therefore also be assessed by examining
customer satisfaction. Nevertheless, we identified only few papers
(20, 14\%) which use qualitative means, and even fewer papers (7,
5\%) in which quantitative means are described to determine a change
in customer satisfaction (see Table~\ref{tab:Success-Indicators-(Continued)}).
Although measuring customer satisfaction by a questionnaire can provide
a more complete view on software quality, it is an intrusive measurement
that needs the involvement and cooperation of the customer~\cite{mccoll-kennedy_measuring_2000}.
On the other hand, quantitative measurements as the number of customer
reported failures need to be put into relation with other, possibly
unknown, variables in order to be a valid measure for software quality.
A decrease in product sales, an increased knowledge of the customer
on how to circumvent problems or a shift in the user base can all
cause a reduction in reported failures, making the measurement of
software quality from this angle more complex~\cite{mockus_predictors_2005}.

\begin{table}
\caption{ISO-9126-1 Product Quality Attributes\label{tab:Product-Quality-Success-Indicators}}

\centering{}\begin{tabular}{l>{\raggedright}p{1.5in}c}
\toprule 
\multicolumn{1}{c}{Quality attribute} & \multicolumn{1}{c}{Studies} & \multicolumn{1}{>{\centering}p{0.6in}}{Frequency (abs/rel)}\tabularnewline
\midrule
\midrule 
Reliability & \cite{prechelt_experiment_2001,roan_pie_1999,hollenbach_portrait_2002,daskalantonakis_practical_1992,henry_quantitative_1995,basili_sels_1995,redzic_six_2006,kuilboer_software_1999,trienekens_targets_2007} & 9/0.47\tabularnewline
\midrule 
Maintainability & \cite{french_applying_1995,hayes_observe-mine-adopt_2003,jarvinen_integrating_1999,kuilboer_software_1999,nelson_evaluating_2002,suardi_how_2004} & 6/0.32\tabularnewline
\midrule 
Reusability & \cite{pfleeger_maturity_1995,basili_sels_1995,calio_software_2000,shah_crafting_1992,nelson_evaluating_2002,tischer_why_2007} & 6/0.32\tabularnewline
\midrule 
Usability & \cite{kuilboer_software_1999,shah_crafting_1992} & 2/0.10\tabularnewline
\midrule 
Portability & \cite{kuilboer_software_1999,shen_measurement_2007} & 2/0.10\tabularnewline
\midrule 
Efficiency & \cite{kuilboer_software_1999} & 1/0.05\tabularnewline
\midrule 
Functionality & \cite{kuilboer_software_1999} & 1/0.05\tabularnewline
\bottomrule
\end{tabular}%
\end{table}

\textbf{Estimation accuracy:} In Table~\ref{tab:Estimation-Accuracy-Success-Indicators}
the success indicators for estimation accuracy are shown. It is interesting
that \emph{estimating} quality seems very uncommon although the improvement
of quality is one of the main interests of SPI initiatives~\cite{gray_limitations_1998,kitchenham_software_1996},
where quality is found to be the most measured success indicator (Table~\ref{tab:Success-indicators}).
The identified quality estimation metric instances cover process quality,
e.\,g.\ actual/estimated number of Quality Assurance reviews~(\cite{mcgarry_attaining_2002})
and actual/estimated number of defects removed per development phase~(\cite{seshagiri_walking_2003}).
Quality estimation metrics should be given equal importance as the
other estimation metrics as they can be used to assess the stability
of the software process. On the other hand, “Schedule” (37, 25\%)
and “Cost and Effort” (34, 24\%) represent the bulk of the estimation
accuracy measures. These two factors may be presumed as important
constraints during the project planning \cite{kerzner_project_2009}
and are therefore preferably selected for estimation.

\begin{table}
\caption{Estimation Accuracy Success Indicators \label{tab:Estimation-Accuracy-Success-Indicators}}

\centering{}\begin{tabular}{l>{\raggedright}p{1.6in}>{\centering}p{0.7in}}
\toprule 
\multicolumn{1}{>{\centering}p{0.45in}}{Success Indicator} & \multicolumn{1}{>{\centering}p{1.6in}}{Studies} & \multicolumn{1}{>{\centering}p{0.6in}}{Frequency (abs/rel)}\tabularnewline
\midrule
\midrule 
Schedule & \cite{regnell_market-driven_1998,escala_metric_1998,sommerville_empirical_2005,roan_pie_1999,hollenbach_portrait_2002,french_applying_1995,daskalantonakis_practical_1992,herbsleb_systematic_1996,paulish_case_1994,iversen_cultivation_2003,achatz_industrial_2003,kautz_making_1999,wang_measuring_2005,grable_metrics_1999,iversen_problems_2006,dion_process_1993,kuntzmann-combelles_quantitative_1995,hashmi_quantitative_2008,gorschek_requirements_2008,ferreira_roi_2008,wohlwend_schlumbergers_1994,kuilboer_software_1999,johnson_software_1994,moreau_software_2003,van_genuchten_using_2001,ahmed_results_2004,pracchia_theav-8b_2004,ebert_impacts_2007,davis_using_2004,hwang_study_2005,shah_crafting_1992,von_wangenheim_experiences_2006,zettell_lipe:_2001,ebert_understanding_2006,el_emam_validating_2000,seshagiri_walking_2003,xu_research_2006} & 37/0.66\tabularnewline
\midrule 
Cost & \cite{roan_pie_1999,hollenbach_portrait_2002,ferreira_applying_2007,herbsleb_systematic_1996,iversen_cultivation_2003,achatz_industrial_2003,iversen_problems_2006,dion_process_1993,gorschek_requirements_2008,ferreira_roi_2008,haley_software_1996,kuilboer_software_1999,johnson_software_1994,moreau_software_2003,mcgarry_what_2002,hwang_study_2005,shah_crafting_1992,el_emam_validating_2000} & 18/0.32\tabularnewline
\midrule 
Effort  & \cite{escala_metric_1998,sakamoto_analysis_1996,damian_empirical_2006,prechelt_experiment_2001,daskalantonakis_practical_1992,cuevas_impact_2007,abrahamsson_personal_2002,henry_quantitative_1995,damian_requirements_2005,calio_software_2000,ebert_technical_1999,abrahamsson_personal_2002-1,davis_using_2004,hwang_study_2005,zettell_lipe:_2001,mcgarry_measuring_1998,xu_research_2006} & 17/0.30\tabularnewline
\midrule 
Size & \cite{prechelt_experiment_2001,taek_lee_cost_2008,jarvinen_establishing_1999,cuevas_impact_2007,anda_improving_2002,grable_metrics_1999,abrahamsson_personal_2002,ebert_technical_1999,davis_using_2004,xu_research_2006} & 10/0.18\tabularnewline
\midrule 
Others & \cite{kuntzmann-combelles_quantitative_1995,hwang_study_2005,seshagiri_walking_2003} & 3/0.05\tabularnewline
\midrule 
Productivity & \cite{prechelt_experiment_2001,seshagiri_walking_2003} & 2/0.04\tabularnewline
\bottomrule
\end{tabular}%
\end{table}

\textbf{Validity of measurements:} Overall we extracted an overwhelming
list of metric instances from the publications. However, many of the
metric instances are actually measuring the same attribute but in
different measurement units, e.\,g.\ defect density which is measured
by taking the number of defects over size, where size can be expressed
in either LOC, FP, etc. Even more interesting is that the definition
of basic measures deviates considerably. For the success indicator
“Productivity” there are examples where the metric was defined as
the ratio of effort over size~(\cite{pfleeger_maturity_1995,mcgarry_measuring_1998}),
and reversely, as the ratio of size over effort~(\cite{wohlwend_schlumbergers_1994,haley_software_1996}).
Another example can be found for the metric \textquotedbl{}Defect
Density\textquotedbl{}, that is interpreted as \textquotedbl{}Process
Quality\textquotedbl{}~(\cite{wohlin_understanding_1998}) but classified
as \textquotedbl{}Defect\textquotedbl{} in~\cite{rooijmans_software_1996,sargut_utilization_2006}. 

A potential reason for these inconsistencies can be the lack of a
by researchers and practitioners acknowledged reference terminology
for software measurement~\cite{garcia_towards_2006}. Imprecise terminology
can lead to inadequate assessment, comparison and reporting of measurement
results and impede learning~\cite{grimstad_software_2006} and therefore
improvement. Besides the lack of agreement on measurement terminology
and concepts, there exist doubts on the validity of certain measures.
The poor definition of measures leads to broad margins of interpretation
as, for example, shown by Kaner and Bond~\cite{kaner_software_2004}
for the reliability metric mean time to failure (MTTF). As pointed
out by Carbone et al.~\cite{carbone_comparison_2008} it is necessary
to understand better the abstract concepts behind the measured quantities
and to construct precise operational definitions in order to improve
the validity of measurements.

\subsection{Identified measurement perspectives in the evaluation of SPI initiatives
(RQ3)\label{sub:What-measurement-perspectives}}

\subsubsection{Results}

\begin{figure}
\begin{centering}
\includegraphics[width=21pc]{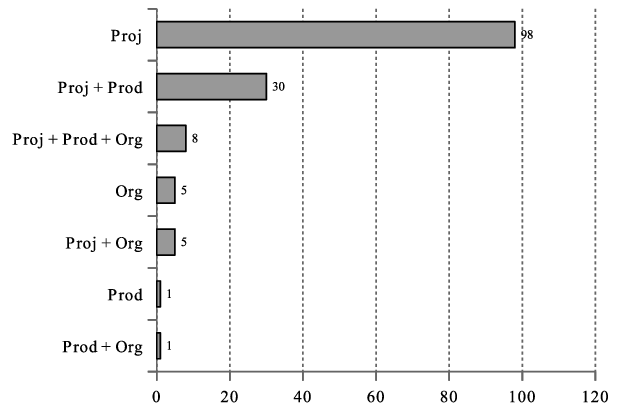}
\par\end{centering}

\caption{Measurement perspective\label{fig:Measurement-perspective}}
\end{figure}

The purpose of this research question was to assess from which measurement
perspective (project, product or organization) SPI initiatives are
evaluated (see Section~\ref{sub:Measurement-perspective} for the
definition of the perspectives). Figure~\ref{fig:Measurement-perspective}
shows the frequencies of the identified measurement perspectives.
The {}``Project'' perspective (98, 66\%) represents the majority,
followed by the {}``Project and Product'' perspective (30, 20\%)
and the {}``Project, Product and Organization'' perspective (8,
5\%). These numbers show that measurement and evaluation at the project
level is the most common approach to assess SPI initiatives

The SPI initiatives and the corresponding measurement perspectives
are mapped in Table~\ref{tab:Measurment-perspectives-established}
and Table~\ref{tab:Measurement-perspectives-variations} respectively.

\begin{table*}
\caption{Measurement perspectives identified in established frameworks\label{tab:Measurment-perspectives-established}}

{\renewcommand{\arraystretch}{1.5}
\begin{tabular}{l@{\extracolsep{8pt}}l@{\vline}p{2.1in}@{\vline\extracolsep{11pt}}p{0.3in}@{\vline}p{0.3in}@{\vline}p{1in}@{\vline}p{0.3in}@{\vline}p{0.3in}@{\vline}p{0.6in}}
\hline
ID & SPI initiatives
&\multicolumn{1}{l@{\vline}}{\begin{sideways}Project (Prj)\end{sideways}}
&\multicolumn{1}{l@{\vline}}{\begin{sideways}Product (Prd)\end{sideways}}
&\multicolumn{1}{l@{\vline}}{\begin{sideways}Organization (Org)\end{sideways}}
&\multicolumn{1}{l@{\vline}}{\begin{sideways}Prj \& Prd\end{sideways}}
&\multicolumn{1}{l@{\vline}}{\begin{sideways}Prj \& Org\end{sideways}}
&\multicolumn{1}{l@{\vline}}{\begin{sideways}Prd \& Org\end{sideways}}
&\multicolumn{1}{l}{\begin{sideways}Prj \& Prd \& Org\end{sideways}}

\tabularnewline \hline \hline  A & CMM &\cite{sakamoto_analysis_1996,dion_elements_1992,schneidewind_measuring_1999,grable_metrics_1999,buchman_software_1996,haley_software_1996,rooijmans_software_1996,ebert_technical_1999,ebert_quest_1998,schalken_using_2006,sargut_utilization_2006,mcgarry_what_2002,diaz_how_1997,mcgarry_measuring_1998,bellini_cmm_2006}&-&\cite{hyde_intangible_2004}&\cite{mckeown_evaluation_2000,achatz_industrial_2003,debou_linking_2000,wohlwend_schlumbergers_1994,johnson_software_1994,batista_spi_2000,trienekens_targets_2007}&\cite{iversen_problems_2006}&-&\cite{herbsleb_systematic_1996,paulish_case_1994,krasner_lessons_1996}

\tabularnewline \hline B & CMMI &\cite{hollenbach_portrait_2002,xu_research_2006}&-&-&\cite{taneike_quality_2008}&-&-&-

\tabularnewline \hline C & SPICE &\cite{hwang_study_2005,el_emam_validating_2000}&-&-&-&-&-&-

\tabularnewline \hline D & PSP &\cite{prechelt_experiment_2001,abrahamsson_personal_2002,abrahamsson_personal_2002-1,wohlin_understanding_1998,humphrey_using_1996}&-&-&-&\cite{taek_lee_cost_2008}&-&-

\tabularnewline \hline E & TSP &\cite{davis_using_2004}&-&-&\cite{seshagiri_walking_2003}&-&-&-

\tabularnewline \hline F & Six-Sigma &\cite{macke_optimized_2008,galinac_software_2007}&-&-&\cite{redzic_six_2006,xiaosong_research_2008}&-&-&-

\tabularnewline \hline G & QIP &\cite{basili_sels_1995}&-&-&-&-&-&-

\tabularnewline \hline H & TQM &-&-&-&\cite{shah_crafting_1992}&-&-&-

\tabularnewline \hline I & IDEAL &\cite{lee_system_2005}&-&-&-&-&-&-

\tabularnewline \hline J & PDCA &\cite{kihara_quality_1998}&-&-&-&-&-&-

\tabularnewline \hline
\multicolumn{2}{c@{\vline}}{Frequency}&30
&0
&1
&12
&2
&0
&3
\tabularnewline \hline
\end{tabular}}%
\end{table*}

\begin{table*}
\caption{Measurement perspectives identified in framework variations, practices
and tools initiatives \label{tab:Measurement-perspectives-variations}}

{\renewcommand{\arraystretch}{1.5}
\begin{tabular}{l@{\extracolsep{8pt}}p{1in}@{\vline\extracolsep{11pt}}p{2.1in}@{\vline}p{0.3in}@{\vline}p{0.3in}@{\vline}p{1in}@{\vline}p{0.3in}@{\vline}p{0.3in}@{\vline}p{0.3in}}
\hline
ID & SPI initiatives
&\multicolumn{1}{l@{\vline}}{\begin{sideways}Project (Prj)\end{sideways}}
&\multicolumn{1}{l@{\vline}}{\begin{sideways}Product (Prd) \end{sideways}}
&\multicolumn{1}{l@{\vline}}{\begin{sideways}Organization (Org)\end{sideways}}
&\multicolumn{1}{l@{\vline}}{\begin{sideways}Prj \& Prd\end{sideways}}
&\multicolumn{1}{l@{\vline}}{\begin{sideways}Prj \& Org\end{sideways}}
&\multicolumn{1}{l@{\vline}}{\begin{sideways}Prd \& Org\end{sideways}}
&\multicolumn{1}{l}{\begin{sideways}Prj \& Prd \& Org\end{sideways}}

\tabularnewline \hline \hline K & Two or more SPI frameworks &\cite{von_konsky_case_2005,sommerville_empirical_2005,ferreira_applying_2007,russ_software_2000,jarvinen_establishing_1999,spork_establishment_2008,jarvinen_integrating_1999,scott_practical_2001,xiaosong_process_2008,kuntzmann-combelles_quantitative_1995,pracchia_theav-8b_2004}&-&\cite{birk_applications_1998,van_solingen_measuring_2004}&\cite{daskalantonakis_practical_1992,murugappan_quality_2000,ferreira_roi_2008,kuilboer_software_1999,moreau_software_2003}&\cite{mcgarry_attaining_2002,dion_process_1993}&-&\cite{miller_sei_2007,savioja_measurement_2007}

\tabularnewline \hline L & Derived SPI framework &\cite{murugappan_blending_2003,cuevas_impact_2007,wang_measuring_2005,cater-steel_process_2006,alagarsamy_knowledge_2007,otoya_experience:_1999}&-&-&-&-&-&\cite{golubic_influence_2005}

\tabularnewline \hline M & Own SPI framework &\cite{weiss_goal-oriented_2002,tanaka_improvement_1995,trienekens_product_2001,canfora_applying_2006}&-&-&\cite{iversen_cultivation_2003,von_wangenheim_experiences_2006}&-&-&-

\tabularnewline \hline N & Limited framework &\cite{visaggio_assessing_2006,lazic_cost_2008,hayes_observe-mine-adopt_2003,karlstraem_minimal_2005,suardi_how_2004,zettell_lipe:_2001}&-&-&\cite{regnell_market-driven_1998}&-&\cite{list_holistic_2005}&-

\tabularnewline \hline \hline O & SPI framework \& Practice &\cite{lee_development_2005,pfleeger_maturity_1995,hashmi_quantitative_2008,nelson_structural_2001,tischer_why_2007}&-&-&\cite{hollenbach_combining_1997,lane_integrating_1997}&-&-&-

\tabularnewline \hline P & SPI framework \& Tool &-&-&-&\cite{escala_metric_1998}&-&-&-

\tabularnewline \hline Q & Practices &\cite{salo_iterative_2007,french_applying_1995,ramil_defining_2000,winkler_early_2007,ajila_empirical_2007,leung_improving_1998,ebert_improving_2001,kautz_making_1999,haugh_never_1991,he_pbr_2006,damian_requirements_2005,biffl_software_2000,ahmed_results_2004,nishiyama_technology_2000,ebert_impacts_2007,van_genuchten_using_2001,nelson_evaluating_2002,haest_evaluation_2000,damm_results_2006,ebert_understanding_2006}&-&\cite{borjesson_improve_2006}&\cite{damian_empirical_2006,baldassarre_industrial_2005,morad_conventional_2005,gou_quantitatively_2008,downey_using_2008,nolan_learning_1999}&\cite{shen_measurement_2007}&-&\cite{roan_pie_1999}

\tabularnewline \hline R & Tool &\cite{visconti_measurement-based_2000,andrade_creation_1999,giraudo_designing_2003,ebert_e-r&d_2002,anda_improving_2002,bruckhaus_impact_1996,haessler_significant_2006}&\cite{momoh_release_2006}&-&\cite{titze_improvement_2000}&-&-&-

\tabularnewline \hline S & Practices \& Tool &\cite{larsen_introducing_1998,dick_lessons_1997,calio_software_2000,winokur_measuring_1998}&-&-&-&-&-&-

\tabularnewline \hline T & Not stated &\cite{freimut_industrial_2005,seacord_measuring_2003,valett_practical_1997,henry_quantitative_1995,el_emam_does_1996}&-&\cite{harrison_technology_1999}&-&-&-&\cite{gorschek_requirements_2008}

\tabularnewline \hline 
\multicolumn{2}{c@{\vline}}{Frequency}
&68
&1
&4
&18
&3
&1
&5
\tabularnewline \hline
\end{tabular}}%
\end{table*}

We identified the organizational measurement perspective mostly in
studies with a CMM-based initiative (row A in Table~\ref{tab:Measurment-perspectives-established}).
We did not identify any study with the product perspective alone within
the established SPI framework category; however rows A, B, E, F and
G in Table~\ref{tab:Measurment-perspectives-established} show that
it is common to combine the project and product perspectives.

\subsubsection{Analysis and Discussion\label{sub:Analysis-and-Discussion-Perspective}}

A considerable amount (98, 66\%) of the total 148 papers in this review
reported only measurements for the project perspective. This indicates
that the measurement perspective to evaluate the SPI initiatives’
outcome is strongly biased towards the project perspective. The dominance
of project perspective and the very low number of organization perspective
may indicate a potential problem to communicate the evaluation results
of the SPI initiatives to all the organization's stakeholders, assuming
that they have different information needs. On the other hand, it
can be argued that measuring the project is easier as probably less
confounding factors are involved~\cite{gorschek_requirements_2008}.

At the corporate level, business benefits realized by the improvement
initiative need to be visible, whereas the initiatives' impact on
a certain project is of more relevance for the involved developers,
project or product managers~\cite{abrahamsson_measuring_2000}. Hence,
it may be beneficial to consider and assess information quality of
software measurements in terms of their fitness of purpose~\cite{berry_assessment_2004}.

It can also be observed that, whenever the product perspective is
considered it is often accompanied by the project perspective. The
combination of these measurement perspectives seems reasonable, especially
when considering the project success definition by Baccarini~\cite{baccarini_logical_1999}:
overall project success is the combination of project management success
and project product success.

Relying exclusively on the project perspective can raise the difficulty
to span the evaluation over several projects, thus not only focusing
on attaining goals of a single project~\cite{gorschek_requirements_2008}.
For example, Babar and Gorton~\cite{babar_software_2009} have observed
in a survey among practitioners that software architecture reviews
are often performed in an ad-hoc manner, without a dedicated role
or team responsible for the review. As such, this initiative may be
beneficial for the current project, but fail to provide the expected
financial benefits in the long-term~\cite{babar_software_2009}.
That would however stay unobserved if the improvement initiative is
only evaluated from the project perspective. It is therefore important
to assess the effect of SPI initiatives from perspectives beyond the
project, that is, consider also the impact on the product and the
organization~\cite{gorschek_requirements_2008}.

Looking at Table~\ref{tab:Measurment-perspectives-established} and
rows K to N in Table~\ref{tab:Measurement-perspectives-variations},
it can be seen that 77 out of 91 (85\%) initiatives that are supported
by a framework are evaluated from the project and/or product perspective.
This indicates a discrepancy of the initiatives aim, i.\,e.\ to
establish an organization-wide improvement (e.\,g.\ at CMM level
3 the improvement is extended to organizational issues~\cite{fitzgerald_longitudinal_1999}),
and how the achievement of this aim is assessed. From the indications
gathered in this review, the organizational measurement perspective
is the least reported one.

SPI initiatives that involve Six Sigma are mostly focused on the {}``Project''
and {}``Project \& Product'' perspective. In 9 out of 10 studies~(\cite{macke_optimized_2008,redzic_six_2006,galinac_software_2007,xiaosong_research_2008}
from Table~\ref{tab:Measurment-perspectives-established} and~\cite{daskalantonakis_practical_1992,murugappan_blending_2003,xiaosong_process_2008,murugappan_quality_2000,hashmi_quantitative_2008}
from Table~\ref{tab:Measurement-perspectives-variations}) these
perspectives are considered while only~\cite{golubic_influence_2005}
from Table~\ref{tab:Measurement-perspectives-variations} covers
the organizational measurement perspective. This could be ascribed
to the emphasis given by Six Sigma on product quality~\cite{biehl_six_2004}
and the implied focus on evaluating the impact on the project and
on the produced goods.

Finally, if we look at the measurement perspectives identified in
the tools and practices category (Table~\ref{tab:Measurement-perspectives-variations},
rows Q, R and S), we can identify some interesting patterns. Only~\cite{borjesson_improve_2006,roan_pie_1999,shen_measurement_2007}
consider the organizational measurement perspective. In particular,
SPI initiatives in the “Tools” and “Practices~+~Tools” categories
do not consider the organization perspective in the measurement. A
potential explanation can be that tools and practices are mostly applied
on project or product levels and not on the organization level. For
the “Practice” category, the most prominent measurement perspective
is the project perspective. The reason is that these initiatives are
mostly addressing the project level. The introduction of a tool as
an SPI initiative can however have far-reaching consequences, that
is for the project~\cite{flynn_is_1995,bruckhaus_impact_1996}, but
also both for the product quality~\cite{kitchenham_case_1995,jarzabek_case_1998,low_software_1999,patnayakuni_development_2002}
and the organization~\cite{humphrey_case_1991,orlikowski_case_1993,premkumar_adoption_1995,sharma_case_2000}.

\subsection{Confounding factors in evaluating SPI initiatives (RQ4) \label{sub:Confounding-factors}}

\subsubsection{Result}

The purpose of this research question was to determine which confounding
factors (see Section~\ref{sub:Confounding-factors-data-extraction})
need to be taken into consideration when evaluating SPI initiatives.
As Table~\ref{tab:Identified-confounding-factors} shows, we could
identify only a few hints regarding these factors. This might indicate
that confounding factors are seldom explicitly taken into consideration
when evaluating process improvement. 

\begin{table*}
\caption{Identified confounding factors\label{tab:Identified-confounding-factors}}

\centering{}\begin{tabular}{c>{\raggedright}p{3.5in}>{\raggedright}p{2.8in}}
\toprule 
Study & \multicolumn{1}{c}{Confounding factors} & \multicolumn{1}{c}{Proposed solutions proposed in the study}\tabularnewline
\midrule
\midrule 
{\footnotesize \cite{karlstraem_minimal_2005}} & {\footnotesize When choosing projects for evaluation, “it is impossible
to find identical projects as they are always of differing size and
nature”.} & {\footnotesize Selection of similar projects in size and nature for
evaluation.}\tabularnewline
\midrule 
{\footnotesize \cite{hollenbach_portrait_2002}} & {\footnotesize Development phase, Measurement unit, Data collection
process} & {\footnotesize Group projects according to project categories and
evaluate categories individually. }\tabularnewline
\midrule 
{\footnotesize \cite{hwang_study_2005}} & {\footnotesize In consecutive assessments, projects may have different
application domains and scope.} & {\footnotesize When comparing two evaluation results, select projects
with similar application domain and scope.}\tabularnewline
\midrule 
{\footnotesize \cite{paulish_case_1994}} & {\footnotesize Environmental influences like staff size and turnover,
capability maturity level, staff morale.} & {\footnotesize Collect those environmental data (influences) which
help to identify and understand influences on performance.}\tabularnewline
\midrule 
{\footnotesize \cite{taek_lee_cost_2008}} & {\footnotesize The result of measuring the impact of personal software
process training depends on the length of project time being measured
and number of sample data used to measure improvement.} & {\footnotesize No solution provided.}\tabularnewline
\midrule 
{\footnotesize \cite{giraudo_designing_2003}} & {\footnotesize The authors identify seven state variables which can
influence the result of their study: programming and testing experience
of the developers, application domain of the tested component, functional
area of the classes involved in the tested component, familiarity
of the developers with other tools, scale of the project, size of
the project team, and number of iterations previously completed.} & {\footnotesize Besides the statement that these variables have to
be taken into consideration when interpreting the result of their
study, no solution is provided.}\tabularnewline
\midrule 
{\footnotesize \cite{diaz_how_1997}} & {\footnotesize Project domain, Project size, Technology changes, code
reuse} & {\footnotesize Project domain - Select similar projects for cycle-time
baselining.}{\footnotesize \par}

{\footnotesize Project size - Normalize size to \textquotedbl{}assembly-equivalent
lines of code\textquotedbl{}.}\tabularnewline
\midrule 
{\footnotesize \cite{ebert_improving_2001}} & {\footnotesize The authors mention uncontrolled independent variables
and the Hawthorne effect \cite{basili_methodology_1984,campbell_hawthorne_1995}.} & {\footnotesize Evaluated projects are grouped according to potential
confounding factors (cultural differences, skill background of employees)
in non-overlapping sets. }\tabularnewline
\midrule 
{\footnotesize \cite{schalken_using_2006}} & {\footnotesize Projects of different size and from different departments.} & {\footnotesize Linear regression models and hierarchical linear models.}\tabularnewline
\midrule 
{\footnotesize \cite{damian_empirical_2006}} & {\footnotesize Organizational changes (management), product maturity,
process changes unrelated to the evaluated improvement initiative,
and the Hawthorne effect.} & {\footnotesize No solution provided except a reasoning why these factors
are minor threats to the internal validity of the study.}\tabularnewline
\midrule 
{\footnotesize \cite{lane_integrating_1997}} & {\footnotesize Staff size, staff training / learning curve, fixed
(\textquotedbl{}overhead\textquotedbl{}) costs as program management,
and configuration management and regression testing for multi-platform
development.} & {\footnotesize Staff size - Production rates normalized to staff size. }{\footnotesize \par}

{\footnotesize Fixed (\textquotedbl{}overhead\textquotedbl{}) costs
– these cost need to be considered in cost reduction improvement.}\tabularnewline
\midrule 
{\footnotesize \cite{larsen_introducing_1998}} & {\footnotesize Changes in the environment that might influence the
experiment results: company restructuring, change of development platform,
changes in product release frequency} & {\footnotesize No solution provided.}\tabularnewline
\midrule 
{\footnotesize \cite{grable_metrics_1999}} & {\footnotesize Project nature (sustainment or new development), manual
data collection, different programming languages, and employee education
an experience level} & {\footnotesize Project nature – Group projects according to project
categories and evaluate the categories individually. }\tabularnewline
\midrule 
{\footnotesize \cite{iversen_problems_2006}} & {\footnotesize {}``Conflicts about measurement goals can often influence
perceptions of success or failure on SPI initiatives”.} & {\footnotesize No solution provided.}\tabularnewline
\midrule 
{\footnotesize \cite{trienekens_product_2001}} & {\footnotesize The authors state that measuring the effect of a specific
process action on product quality is possible. However, the missing
knowledge on relationships between process actions and product quality
makes the measurement unreliable and therefore it cannot be generalized
to all situations.} & {\footnotesize No solution provided.}\tabularnewline
\midrule 
{\footnotesize \cite{gou_quantitatively_2008}} & {\footnotesize The authors state that the availability and quality
of historical data can affect the result of applying their method
of defect-related measurement (BiDefect).} & {\footnotesize Data from a stable process is required if no high quality
historical data is available.}\tabularnewline
\midrule 
{\footnotesize \cite{ferreira_roi_2008}} & {\footnotesize Several factors that can influence the productivity
values such as language, project size, tools and technical issues.} & {\footnotesize Measuring projects that use same language, tools, development
environment and normalizing the productivity by size (function points)
can help to reduce the influence of those factors.}\tabularnewline
\midrule 
{\footnotesize \cite{van_genuchten_using_2001}} & {\footnotesize {}``The preparation rate is known to be a main independent
variable for inspection quality.''} & {\footnotesize Measure preparation rates in software inspections and
take them into account when evaluating the efficiency and effectiveness
of software inspections.}\tabularnewline
\midrule 
{\footnotesize \cite{downey_using_2008}} & {\footnotesize Development team / Test (QA) team, technology, customer} & {\footnotesize Staffing, technology and used platform in all projects
is similar. Customer is the same organizational division.}\tabularnewline
\bottomrule
\end{tabular}%
\end{table*}

\subsubsection{Analysis and Discussion}

From the results presented above we can identify several issues regarding
confounding factors and their role in evaluating SPI initiatives.
The first is that we could only identify 19 studies (out of 148) which
discuss potential validity problems when evaluating SPI initiatives.
It is therefore difficult to generalize assumptions or to relate a
finding to a certain evaluation strategy. Second, the authors of the
publications seldom use the term “confounding factor” or “confounding
variable''; often we had to interpret the descriptions of study designs,
executions and results to discover if the authors considered confounding
factors. We identified several synonyms instead: “influencing factors”~\cite{ferreira_roi_2008},
“influences”~\cite{grable_metrics_1999}, “state variables”~\cite{giraudo_designing_2003},
“uncontrolled independent variables”~\cite{ebert_improving_2001}
and “environmental influences”~\cite{paulish_case_1994,larsen_introducing_1998}. 

What can be learned from the identified studies is that the identification,
characterization and control of confounding factors is a challenging
endeavor. In~\cite{taek_lee_cost_2008,larsen_introducing_1998,iversen_problems_2006,trienekens_product_2001}
they are described in an abstract and general way without discussing
remedies to overcome them. The authors in~\cite{trienekens_product_2001}
pointed out that it is possible to measure product quality improvement
effected by specific process actions. They also cautioned that it
is necessary to study the conditions under which the relationship
between process action and improvement are observed in order to increase
the knowledge on these relationships. Unfortunately in many cases
the context, in which the improvement is evaluated, is described unsatisfactorily
(see Section~\ref{sub:Study-context}), and an identification of
confounding factors is therefore aggravated. 

Generally, the effect of confounding factors on the dependent variable
can be controlled by designing the study appropriately, e.\,g.\ by
a random allocation of the treatment and control groups~\cite{anderson_statistical_1980}.
The fundamental assumption by such a design is that the confounding
variables are equally distributed in each group, i.\,e.\ that the
probability is high that the groups have similar properties. Therefore,
if the distribution of the dependent variable is similar in both the
control and treatment group, it can be concluded that the treatment
has no effect. 

The concept of randomization is also discussed in~\cite{pfleeger_experimentation_1997,kitchenham_preliminary_2002,wohlin_empirical_2003}
in the context of software engineering experiments. Pfleeger~\cite{pfleeger_experimentation_1997},
points out that the major difference between experiments and case
studies is the degree of control. In order to control a potential
confounding variable, the experiment can be designed in such a way
that the experimental units within the distinct groups are homogeneous
(blocking). Additionally, if the number of experimental units is the
same in each group, the design is balanced. 

Unfortunately, random sampling of projects or subjects is seldom an
option in the evaluation of improvement initiatives and therefore
knowing of the existence of potential confounding factors is however
needed in order to be able to apply certain techniques to compensate
confounding effects~\cite{anderson_statistical_1980}. The matching
technique, for example, leads to an evaluation design that satisfies
the ceteris paribus condition by selecting groups with similar properties
with respect to confounding factors~\cite{anderson_statistical_1980}.
By looking at the proposed solutions, several studies apply some sort
of matching, e.\,g.\ by selecting similar projects in terms of size
and application domain, technology or staff size (see~\cite{hollenbach_portrait_2002,ebert_improving_2001,karlstraem_minimal_2005,hwang_study_2005,diaz_how_1997,grable_metrics_1999,ferreira_roi_2008,downey_using_2008}
in Table~\ref{tab:Identified-confounding-factors}). 

There exists no systematic way to identify confounding variables~\cite{pearl_why_1998}
and as shown by the examples above, their identification depends on
the context in which the study is conducted and on the background
knowledge of the researcher. It is therefore difficult to assure that
all confounding variables are eliminated or controlled, since their
determination relies on assumptions and sound logical reasoning. An
interesting discussion about the identification of a confounding factor
can be found in the comments by Evanco~\cite{evanco_comments_2003},
which refers to the validity of the assumption by El Emam et al.\ that
size is a confounding variable for object oriented metrics~\cite{el_emam_confounding_2001}.
El Emam et al.\ demonstrate empirically that class size confounds
the validity of object oriented metrics as indicators of the fault-proneness
of a class. The comments~\cite{evanco_comments_2003}, however, show
that the identification and recognition of certain confounding factors
is still disputed~\cite{bruntink_empirical_2006,zhou_examining_2009}.

\section{Conclusion\label{sec:Conclusion}}

This paper presents a systematic literature review that investigates
how the impact of software process improvement initiatives (as defined
in Section~\ref{sub:SPI-initiative}) is measured and evaluated.
The aim is to identify and characterize the different approaches used
in realistic settings, i.\,e.\ to provide a comprehensive outline
and discussion of evaluation strategies and measurements used in the
field to assess improvement initiatives. The major findings of this
review and their implications for research are:
\begin{itemize}
\item \textbf{Incomplete context descriptions:} Seventy-five out of 148
studies did not or only partially describe the context in which the
study was carried out (see Section~\ref{sub:Study-quality-assessment}).
In the area of process improvement it is however critical to describe
the process change and its environment in order to provide results
which have the potential to be reused or to be transferred into different
settings. Since a considerable body of knowledge on the impact of
improvement initiatives is provided by industry reports (53, 36\%),
a precise and informative context description would be beneficial
for both practitioners and researchers.
\item \textbf{Evaluation validity:} In more than 50\% of the studies in
which improvement initiatives are evaluated, {}``Pre-Post Comparison''
is used individually or in combination with another method (see Section~\ref{sub:Evaluation-strategies-RQ1}).
Considering that confounding factors are rarely discussed (19 out
of 148 studies, see Section~\ref{sub:Confounding-factors}), the
accuracy of the evaluation results can be questioned. The severity
of confounding is even increased by unsatisfactory context descriptions.
A grounded judgment by the reader on the validity of the evaluation
is prohibited by the absence of essential information.
\item \textbf{Measurement validity:} Kaner and Bond~\cite{kaner_software_2004}
illustrated how important it is to define exactly the semantics of
a metric and the pitfalls that arise if it is not commonly agreed
what the metric actually means, i.\,e.\ which attribute it actually
measures. This issue is related with farther reaching questions than
process improvement measurement and evaluation, and concerns fundamental
problems of software measurement validity. Nevertheless, measurement
definition inconsistencies, as shown in Section~\ref{sub:Analysis-and-Discussion-Metrics},
inhibit the process of improvement itself since the comparison and
communication of results is aggravated. The implication for research
is that it is difficult to identify and use the appropriate measures
for improvement evaluation. A better support for defining, selecting
and validating measures could enable a comparable and meaningful evaluation
of SPI initiatives.
\item \textbf{Measurement scope:} The analysis on what is actually measured
during or after an improvement initiative shows a focus on process
and product quality (see Section~\ref{sub:Reported-Metrics}). From
the software \emph{process} improvement perspective this measurement
goal might be adequate and sufficient. It is however crucial to push
the event horizon of improvement measurement beyond the level of projects
(see Section~\ref{sub:What-measurement-perspectives}) in order to
confirm the relatively short-dated measurements at the project or
product level. Since the information needs for the different stakeholders
vary, appropriate improvement indicators need to be implemented. At
the corporate level for example, business benefits realized by projects
which encompass a wider scope than pilot improvement implementations
are of interest.\\
Indicators for these long-term effects can be customer satisfaction,
to assess quality improvement, and return on investment to evaluate
the economic benefits of improvement. The data presented in this review
(see Section~\ref{sub:Analysis-and-Discussion-Metrics}) suggests
that these indicators tend to be less used in the evaluation of process
improvement as other, easier to collect, indicators. The implication
for research is to integrate the success indicators into a faceted
view on process improvement which captures its short- and long-term
impact.
\item \textbf{Confounding factors:} In a majority (129, 87\%) of the reviewed
studies we could not identify a discussion on confounding factors
that might affect the performance of SPI initiatives and thus their
evaluation. Since process improvement affects many aspects of a development
project, its results and effect on the organization, there are many
potential such confounding factors that threaten validity. Even though
study design can often be used to limit the effects it is often not
practical to fully control the studied context. Thus future research
on SPI should always consider and discuss confounding factors. However,
we note that no good conceptual model or framework for such a discussion
is currently available.
\end{itemize}
The results of this review encourage further research on the evaluation
of process improvement, particularly on the conception of structured
guidelines which support practitioners in the endeavor of measuring,
evaluating and communicating the impact of improvement initiatives.

\section*{Acknowledgment}

\addcontentsline{toc}{section}{Acknowledgment} 

The authors thank the anonymous reviewers whose detailed and judicious
comments improved the paper considerably. This work was partially
funded by the Industrial Excellence Center EASE - Embedded Applications
Software Engineering (http://ease.cs.lth.se).

\bibliographystyle{IEEEtran}
\bibliography{Evaluation_and_Measurement_of_Software_Process_Improvement-A_Systematic_Literature_Review}

\begin{IEEEbiography}[{\includegraphics[width=1in,height=1.25in,clip, keepaspectratio]{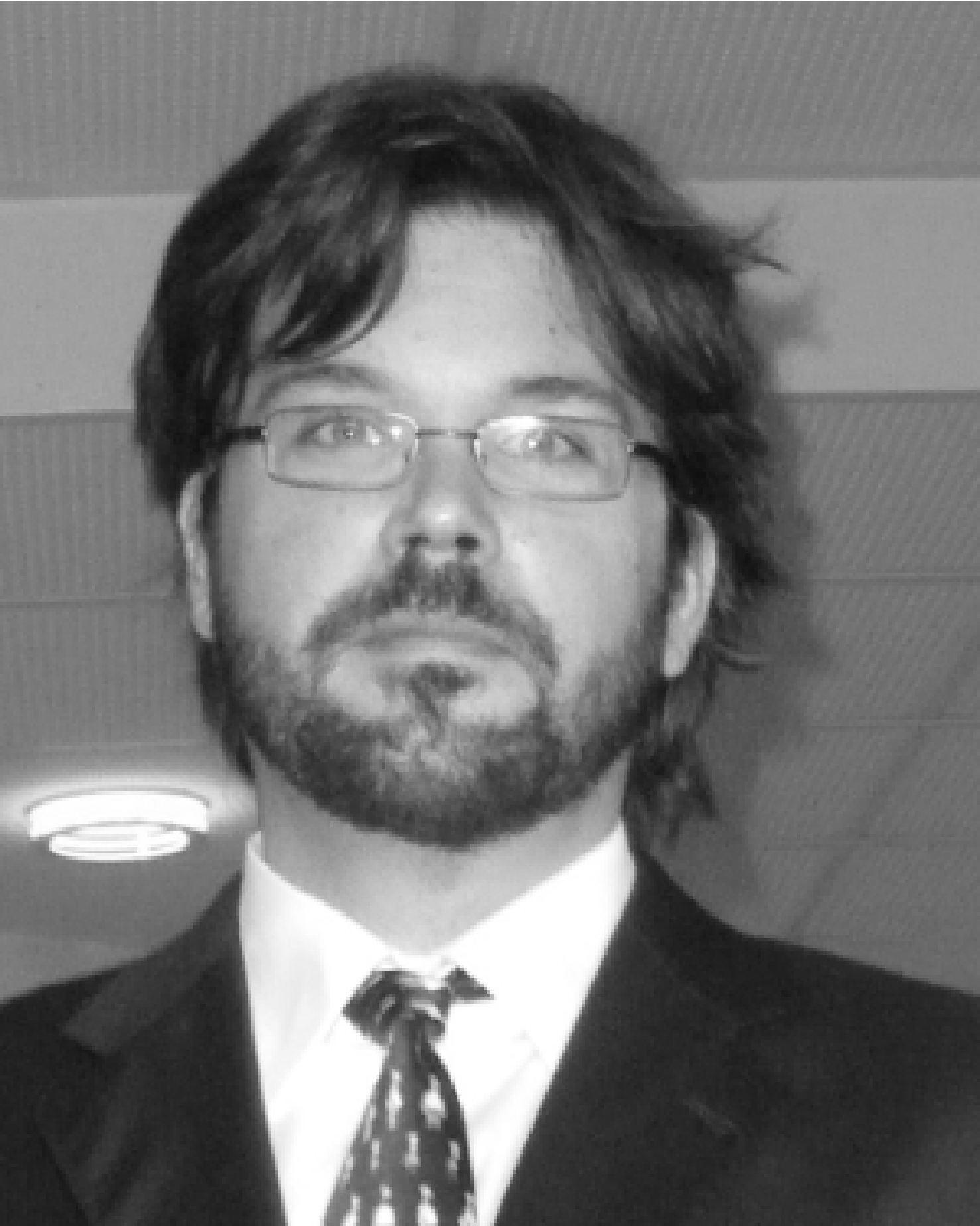}}]{Michael~Unterkalmsteiner} is a PhD student at the Blekinge Institute of Technology (BTH) where he is with the Software Engineering Research Lab. His research interests include software repository mining,  software measurement and testing, process improvement, and requirements engineering. His current research focuses on the co-optimization of requirements engineering and verification \& validation processes. He received the B.\,Sc. degree in applied computer science from the Free University of Bolzano\,/\,Bozen (FUB) in 2007 and is currently completing the M.\,Sc. degree in software engineering at BTH.
\end{IEEEbiography}
\begin{IEEEbiography}[{\includegraphics[width=1in,height=1.25in,clip,keepaspectratio]{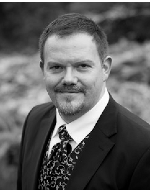}}]{Tony~Gorschek}
is a professor of software engineering at Blekinge Institute of Technology (BTH) with over ten years industrial experience. He also manages his own industry consultancy company, works as a CTO, and serves on several boards in companies developing cutting edge technology and products. His research interests include requirements engineering, technology and product management, process assessment and improvement, quality assurance, and innovation. Contact him at tony.gorschek@bth.se or visit www.gorschek.com.
\end{IEEEbiography}
\begin{IEEEbiography}[{\includegraphics[width=1in,height=1.25in,clip,keepaspectratio]{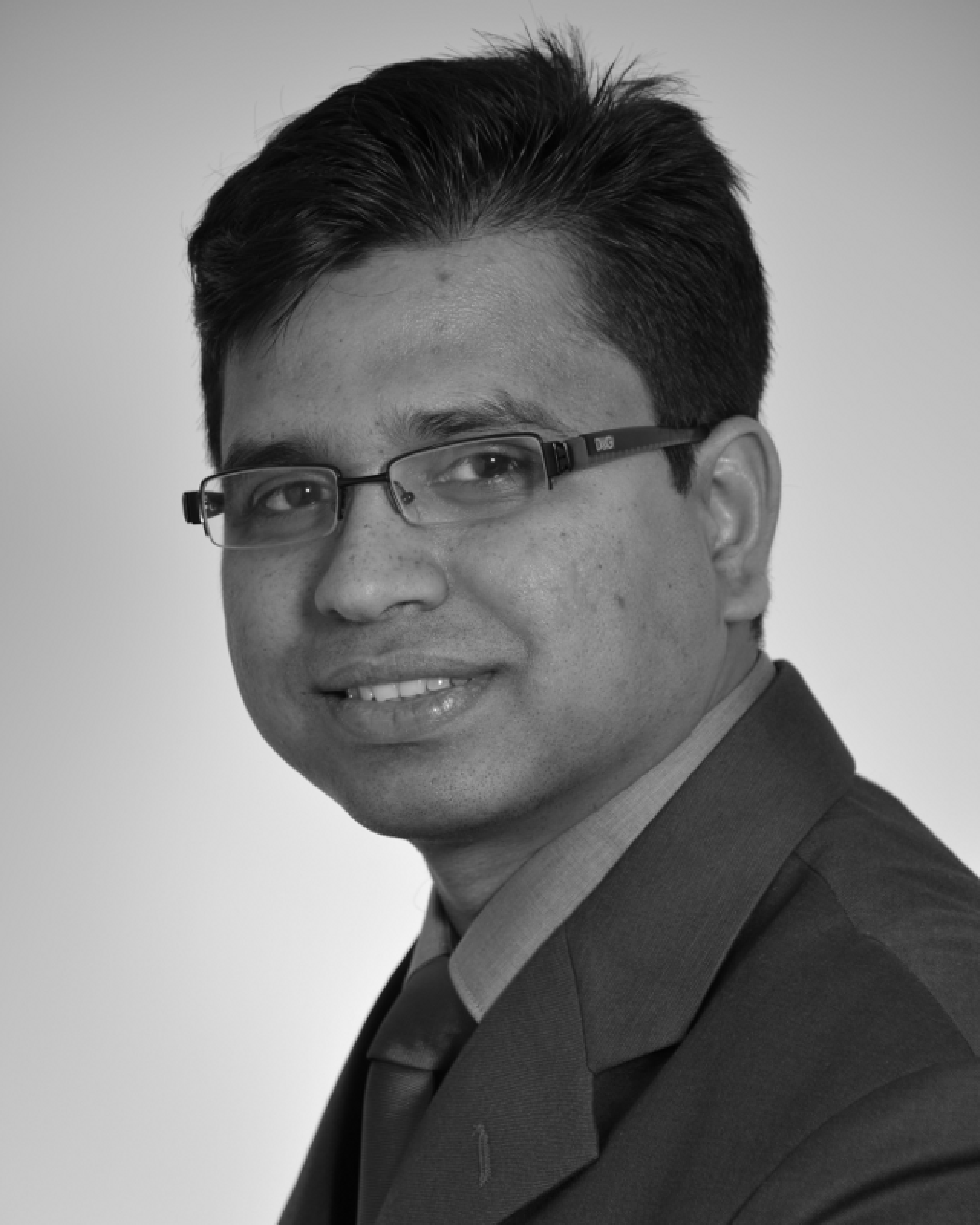}}]{A.\,K.\,M.~Moinul~Islam}
is a researcher at the Technical University of Kaiserslautern, Germany. He is with the Software Engineering: Process and Measurement Research Group. His research interests include global software engineering, software process improvement and evaluation, and empirical software engineering. He received his double master’s degree, M.\,Sc. in Software Engineering, in 2009 jointly from University of Kaiserslautern, Germany and Blekinge Institute of Technology, Sweden within the framework of European Union's Erasmus Mundus Programme. Prior to his master's degree, he worked for 3 years in the IT and Telecommunication industry.
\end{IEEEbiography}
\begin{IEEEbiography}[{\includegraphics[width=1in,height=1.25in,clip,keepaspectratio]{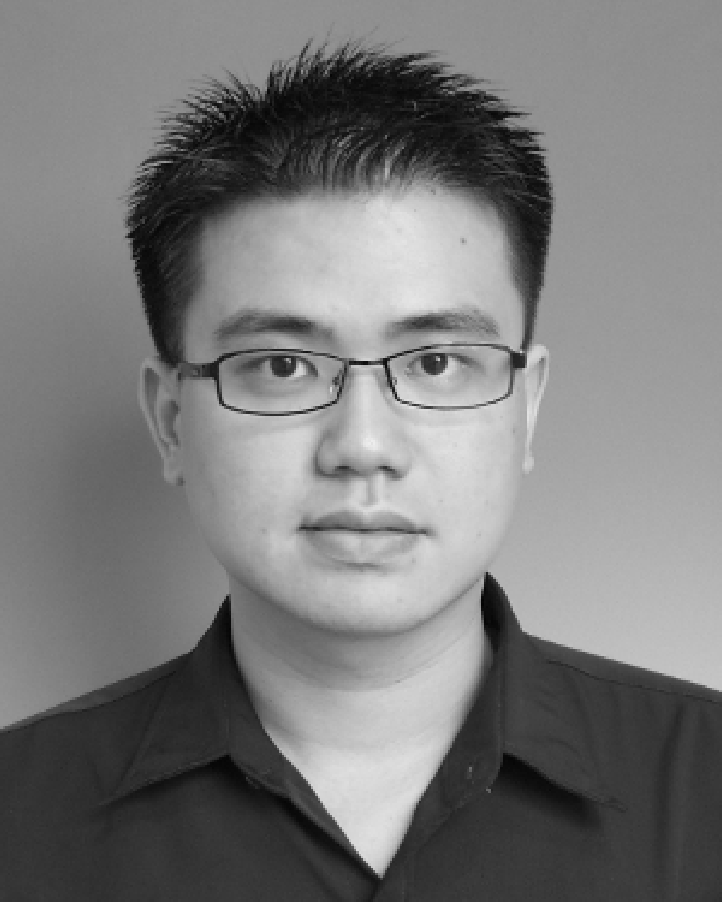}}]{Chow~Kian~Cheng}
is a software engineer at General Electric International Inc. based in Freiburg, Germany. He is responsible for the development of clinical software in the healthcare industry. He holds a joint master degree, M.\,Sc. in Software Engineering, from the Blekinge Institute of Technology, Sweden and the Free University of Bolzano\,/\,Bozen, Italy. Prior to his master degree, he worked for 4 years with Motorola Inc. and Standard Chartered Bank.
\end{IEEEbiography}
\begin{IEEEbiography}[{\includegraphics[width=1in,height=1.25in,clip,keepaspectratio]{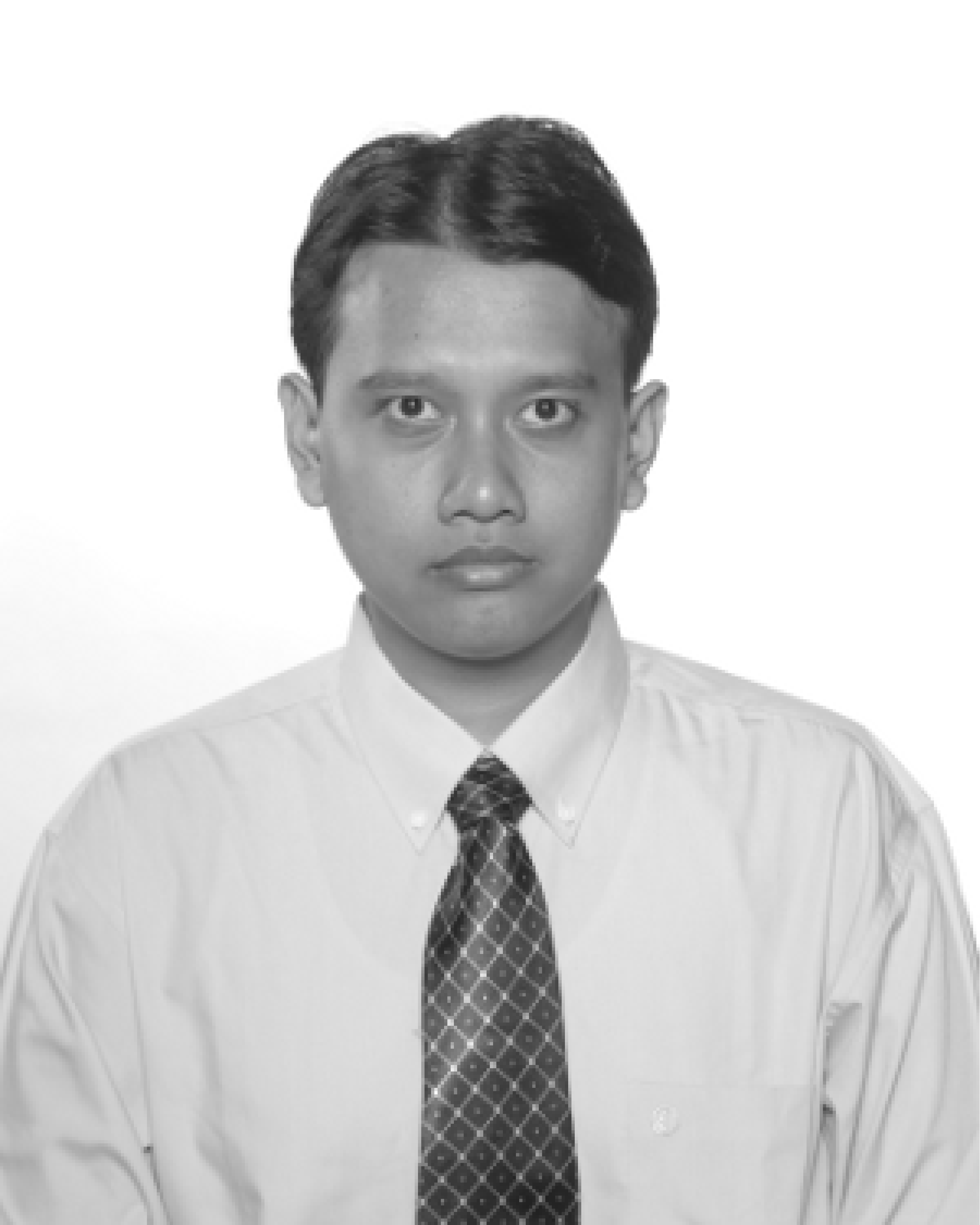}}]{Rahadian~Bayu~Permadi}
received his Bachelor degree in Informatics from Bandung Institute of Technology, Indonesia. He obtained in 2009 the double-degree master in software engineering from the Free University of Bolzano\,/\,Bozen, Italy and the Blekinge Institute of Technology, Sweden. Currently, he is working as a software engineer at Amadeus S.\,A.\,S, France. His interests are software measurements \& process improvement, software architecture and software project management. He was a Java technology researcher in Indonesia before he got awarded with the Erasmus Mundus scholarship for European Master in Software Engineering programme.
\end{IEEEbiography}
\begin{IEEEbiography}[{\includegraphics[width=1in,height=1.25in,clip,keepaspectratio]{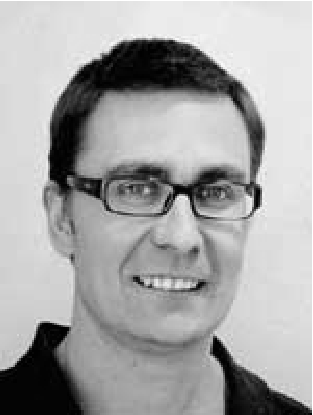}}]{Robert~Feldt}
(M’98) is an associate professor of software engineering at Chalmers University of Technology (CTH) as well as at Blekinge Institute of Technology. He has also worked as an IT and Software consultant for more than 15 years. His research interests include software testing and verification and validation, automated software engineering, requirements engineering, user experience, and human-centered software engineering. Most of the research is conducted in close collaboration with industry partners such as Ericsson, RUAG Space and SAAB Systems. Feldt has a PhD (Tekn.\,Dr.) in software engineering from CTH.
\end{IEEEbiography}
\end{document}